\documentclass[preprint,10pt]{aastex}

\usepackage{amssymb,amsmath,amsfonts}
\usepackage{graphicx}
\usepackage{color}
\usepackage{ulem}             

\newcommand\lam{{\lambda} }
\newcommand\Lam{{\Lambda} }
\newcommand\om{{\omega} }
\newcommand\sig{{\sigma} }
\newcommand\gam{{\gamma} }
\newcommand\Gam{{\Gamma} }
\newcommand\eps{{\varepsilon} }

\newcommand{\NN}{{\mathbb N}}

\newcommand\br{{\bf r} }
\newcommand\bu{{\bf u} }
\newcommand\bE{{\bf E} }
\newcommand\bC{{\bf C} }
\newcommand\brt{\tilde{\bf r} }
\newcommand\ab{{\overline a} }
\newcommand\xb{{\overline x} }
\newcommand\yb{{\overline y} }

\newcommand\Xb{{\overline X} }
\newcommand\Yb{{\overline Y} }
\newcommand\Bb{{\overline B} }
\newcommand\Hb{{\overline H} }
\newcommand\sab{\sqrt{\overline a} }

\newcommand\zetab{{\overline \zeta} }

\newcommand\xt{{\tilde x}}

\newcommand\zt{{\tilde z}}
\newcommand\qt{{\tilde q} }
\newcommand\zetat{{\tilde \zeta} }

\newcommand{\qtext}[1]{\quad \text{#1}\quad}

\newcommand{\norm}[1]{\vert\vert#1 \vert\vert}
\newcommand\cG{{\cal G} }
\newcommand\gO{{\cal O} }
\newcommand\cS{{\cal S} }
\newcommand\cD{{\cal D} }
\newcommand\cF{{\cal F} }

\newcommand\bfp{{\bf p} }
\newcommand\bfq{{\bf q} }

\newcommand{\vect}[3]{\left(\hspace{-0pt} \begin{array}{c}  #1 \\  #2 \\  #3 \end{array}\hspace{-0pt} \right)}

\newcommand{\be}{\begin{equation}}
\newcommand{\ee}{\end{equation}}

\newcommand{\dron}[2]{\frac{\partial#1}{\partial#2}}

\newcommand{\tpr}[1]{{#1}}

\newenvironment{disarray}%
 {\everymath{\displaystyle\everymath{}}\array}%
 {\endarray}

\begin{document}

\title{On the co-orbital motion of two planets in quasi-circular orbits}

\author{Philippe Robutel and  Alexandre Pousse}

\affil{IMCCE, Observatoire de Paris, UPMC, CNRS UMR8028, 77 Av. Denfert-Rochereau, 75014 Paris, France \\
              } 
                      
\date{\today}

\begin{abstract}
We develop an analytical Hamiltonian formalism adapted to the study of the motion of two planets in co-orbital resonance. The Hamiltonian, averaged over one of the planetary mean longitude, is expanded in power series of eccentricities and inclinations. The model, which is valid in the entire co-orbital region, possesses an integrable approximation modeling  the planar and quasi-circular motions.   
First, focusing on the fixed points of this approximation,   we highlight relations linking the eigenvectors of the associated  linearized differential system  and the existence of certain remarkable orbits like  the elliptic Eulerian Lagrangian configurations, the Anti-Lagrange (Giuppone et al., 2010) orbits and some second sort orbits discovered by Poincar\'e.   Then, the variational equation is studied in the vicinity of any  quasi-circular periodic solution. The fundamental frequencies of the trajectory are deduced and possible occurrence of low order resonances are discussed.  Finally,  with the help of the construction of a Birkhoff normal form, we prove that the  elliptic Lagrangian equilateral 
configurations and the Anti-Lagrange orbits bifurcate from the same fixed point $L_4$. 
\end{abstract}




\section{Introduction}
\label{sec:intro}

The co-orbital resonance has been extensively studied for more than one hundred years in the framework of the restricted three-body problem (RTBP). 
In most of the analytical works, the emphasis has been placed on the tadpole orbits, trajectories surrounding one of the two Lagrangian triangular equilibrium points,  since these describe the motion of the Jovian Trojans.  
However, the global topology of the co-orbital resonance has been studied in particular by  \cite{Gar1976a,Gar1978a} and   \cite{Ed1977},  but the interest for the horseshoe orbits which encompass  the three equilibrium points $L_3, L_4$ and $L_5$, remained  academic until the discovery of the Saturnian satellites  Janus and Epimetheus \citep{SmiReFoLa1980,SyPeSmiMo1981}.
In \cite{DeMu1981a}, general properties of the tadpole and horseshoe orbits are described in the quasi-circular case. 
In particular, asymptotic  estimates of the  horseshoe orbits  lifetime  and the relative width of this orbits domain are given.  But the  impossibility to get explicit expressions of the  horseshoe orbits complicated their study, and theoretical works were replaced   by numerical simulations. 
 Thereby, \cite{Chr2000} showed that the region containing the tadpole orbits is not disconnected from the horseshoe one, and that there exist transitions between these two domains. 
Also, a global study of the phase space of the co-orbital resonance was presented in the RTPB by \cite{NeThoFeMo02} using a numerical  averaging of the disturbing function over orbital frequencies. 
Using the same kind of numerical technics,  \cite{GiuBeMiFe2010}  studied the stability regions and families of periodic orbits of two planets locked in the co-orbital resonance.  Besides the  Lagrangian triangular configurations, where the three bodies in Keplerian motion occupy the vertices of an equilateral triangle, these authors found a new family of fixed points  (equilibrium in the reduced average problem, but quasi-periodic with two fundamental frequencies for the non-average problem in inertial reference frame) that they called Anti-Lagrange orbits.
Both Lagrange and Anti-Lagrange families can be seen as a one-parameter family of stable fixed points, parametrized by the eccentricity. 
As shown in the Fig. 7 of \cite{GiuBeMiFe2010}, when the eccentricity is equal to zero, the corresponding configurations of each family seem to merge  in the  well known circular Lagrangian equilateral configuration. Finally,  when the eccentricity increases, the stability regions surrounding these orbits become smaller, the one associated to the Lagrangian configuration being the first to vanish. 

In this paper, we will develop a Hamiltonian formalism adapted to the study of the motion of two planets in co-orbital resonance.  We modify  the methods presented in \cite{LaRo1995} in order to get an analytical expansion of the planetary Hamiltonian averaged over  an orbital period, meaning  averaged over  one of the planetary mean longitudes. This expansion, which is a power series of the eccentricities and inclinations whose coefficients depend on the semi-major axes and on the difference of the planetary mean longitudes,  generalizes the expressions obtained in the RTBP framework by \cite{Morais1999,Morais2001}. 
Moreover, this one containing only  even terms in the eccentricities and inclinations (see Section \ref{sec:averadgedro}), the planar circular motions are conserved along the solutions. These quasi-circular co-orbital motions are modeled by an integrable Hamiltonian depending only on the semi-major axes and the mean longitudes difference. Contrary to the integrable approximations derived by \cite{YoCoSyYo1983} or \cite{Morais1999,Morais2001}, our model
possesses five fixed points. Three are unstable and correspond to the Eulerian collinear configurations denoted by $L_1$, $L_2$ and $L_3$ where the two planets revolve on circles centered at the Sun.  The two others correspond to the  circular Lagrangian equilateral configurations $L_4$ and $L_5$  mentioned above, which are linearly stable if the  planetary masses are small enough \citep{Ga1843}. 

In Section \ref{sec:quadratic}, the linear differential system associated to infinitesimal variations  transversal to the plane containing the quasi-circular orbits will be studied. Only the directions corresponding  to the eccentricities will be considered. First, focusing on the fixed points,   we will highlight relations linking the eigenvectors of the linearized differential system  and the existence of certain remarkable orbits like  the elliptic Eulerian and the Lagrangian configurations, the Anti-Lagrange orbits and some second sort orbits discovered by \cite{Poi1892}.   Then, the variational equation will  be studied in the vicinity of any  quasi-circular periodic solution. The fundamental frequencies of the trajectory will be deduced, and possible occurrence of low order resonances will be discussed.  

Section \ref{sec:LypFam} will be devoted to the construction of a Birkhoff normal form in the neighborhood of $L_4$. Beside the derivation of the fundamental frequencies of any quasi-periodic trajectory lying in this neighborhood, we will prove that the  elliptic Lagrangian equilateral 
configurations and the Anti-Lagrange orbits bifurcate from the same fixed point $L_4$. 
Finally, in  Section \ref{sec:comments}, comments and   various approaches for future works will be presented.

\section{The average Hamiltonian}
\label{sec:averadgedro}
\subsection{Canonical heliocentric coordinates}
We consider two planets of respective masses $m_1$ and $m_2$ orbiting a central body (Sun, or star) of mass $m_0$ dominant with respect to the planetary masses.  
As only co-orbital planets are considered, no planet is permanently farther from the central body than the other, so the heliocentric coordinate system seems to be the most adapted to this situation. Following \cite{LaRo1995}, the Hamiltonian of the three-body problem reads
\be
\begin{split}
H(\brt_j,\br_j) = H_K(\brt_j,\br_j) + H_p(\brt_j,\br_j) \quad \text{with}  \\
 H_K(\brt_j,\br_j) =  \sum _{j\in\, \{1,2\}} \left(
 \frac{\brt_j^2}{2\beta_j} - \frac{\mu_j \beta_j}{\norm{\br_j}} 
    \right)
    \quad \text{and} \\
  H_p(\brt_j,\br_j) = \frac{\brt_1\cdot\brt_2}{m_0} - \cG\frac{m_1m_2}{\norm{\br_1 -\br_2}}, 
     \end{split}
     \label{eq:ham_cart}
\ee
%
where $\br_j$ is the heliocentric position of the planet $j$, $\beta_j = m_0m_j(m_0+m_j)^{-1}$ and $\mu_j = \cG(m_0+m_j)$, $\cG$ being the gravitational constant. The conjugated variable of $\br_j$, denoted by $\brt_j$, is the barycentric linear momentum of the body of index $j$.  In this expression,   $H_K$ corresponds to the unperturbed Keplerian motion of the two planets, more precisely the motion of a mass $\beta_j$ around a fixed center of mass $m_0+m_j$, while  $H_p$ models the gravitational perturbations. If we introduce the small parameter $\eps$ given by
\be
\eps = \text{Max}\left(\frac{m_1}{m_0},\frac{m_2}{m_0}\right),
\ee
one can verify that the Keplerian term of the planetary Hamiltonian is of order $\eps$ and  the other one is of  order $\eps^2$ which justifies a perturbative approach. 

The choice of these canonical heliocentric coordinates $(\brt_j,\br_j)$ may  lead to quite surprising results for quasi-circular motions.  In particular, the famous Lagrangian relative equilibrium,  where the three bodies occupying the vertices of an equilateral triangle animated with an uniform rotation, is described  in terms of elliptical elements by ellipses in rapid rotation. 
More precisely, $\brt_j$ being not collinear to the heliocentric velocity of the planet $j$, the Keplerian motion associated to the unperturbed Hamiltonian $\brt_j^2/(2\beta_j) - \mu_j \beta_j / \norm{\br_j}$  is not represented by a circle described with constant angular velocity, but by a  rapidly precessing ellipse whose eccentricity is proportional to the planetary masses.
 This phenomenon described in the appendix (Section \ref{sec:appendix}) is similar to the question of the definition of elliptical elements for a satellite orbiting an oblate body (see \cite{Gr1981}).
Except this little drawback which occurs only when the considered motion is close to the circular Lagrangian configurations, the canonical heliocentric variables are  particularly \tpr{well suited to the study of } the co-orbital resonances. 

In order to define a canonical coordinate system related to the elliptical elements $(a_j,e_j,I_j,\lam_j,\varpi_j,\Omega_j)$ (respectively the semi-major axis, the eccentricity, the inclination, the mean longitude, the longitude of the pericenter and the longitude of the ascending node of the planet $j$), we start from  Poincar\'e's rectangular variables in complex form $(\lam_j,\Lam_j,x_j,-i\xb_j,y_j,-i\yb_j)$ where $\Lam_j = \beta_j\sqrt{\mu_ja_j}$,
\be
\begin{split}
x_j = \sqrt{\Lam_j}\sqrt{1-\sqrt{1-e_j^2}}\exp(i\varpi_j), \\
y_j = \sqrt{\Lam_j}\sqrt{\sqrt{1-e_j^2}(1-\cos I_j)}\exp(i\Omega_j).
\end{split}
\ee
This coordinate system has the advantage of being regular when the eccentricities and the inclinations tend to zero. 
It is also convenient to use the non-dimensional quantities  $X_j = x_j\sqrt{2/\Lam_j}$ and $Y_j = y_j/\sqrt{2\Lam_j}$ which are equivalent to $e_j\exp(i\varpi_j)$ and $I_j\exp(i\Omega_j)/2$ for quasi-planar and quasi-circular motions. 

As we only consider the planetary motions in the vicinity of the circular planar problem, the Hamiltonian can be expanded in power series of the variables $X_j, Y_j$ and their conjugates in the form
\be
\sum_{k_1,k_2} \left(
\sum_{(\bfp, \bfq)\in \mathbb{N}^8} \Psi_{\bfp, \bfq}^{k_1,k_2}(\Lam_1,\Lam_2)X_1^{p_1}X_2^{p_2}\Xb_1^{\bar p_1}\Xb_2^{\bar p_2}Y_1^{q_1}Y_2^{q_2}\Yb_1^{\bar q_1}\Yb_2^{\bar q_2}
\right)
e^{i(k_1\lam_1+k_2\lam_2)},
\ee
where the  integers occurring in these summations satisfy the relation 
\be
\sum_j (k_j + p_j + q_j - \bar p_j - \bar q_j) = 0, \label{D'Al}
\ee
 known as D'Alembert rule. This relation corresponds to the invariance of the Hamiltonian by rotation, or, which is equivalent, to the fact that the angular momentum of the system is an integral of the motion. Remark that we will not  use this explicit Fourier expansion in this paper, but the D'Alembert rule will play an important role. 

  According to \tpr{\cite{Poincare1905}}, the expression of the angular momentum in Poincar\'e's variables reads
\be
\bC = \sum_j  \br_j \times \brt_j =
 \sum_j  \vect
 {\sqrt2 \Im(y_j)\sqrt{\Lam_j  - \vert x_j\vert^2 - \frac{\vert y_j \vert^2}{2}  }}
 {-\sqrt2 \Re(y_j)\sqrt{\Lam_j  - \vert x_j\vert^2 - \frac{\vert y_j \vert^2}{2}  }}
 {\Lam_j  - \vert x_j\vert^2 - \vert y_j \vert^2}.
\ee

In order  to deal with the  co-orbital resonance, an appropriate canonical coordinate system is\footnote{Other coordinates adapted to the co-orbital resonance have been used by several authors (e.g. \cite{NeThoFeMo02} for the RTBP and \cite{GiuBeMiFe2010} for the planetary problem), but these systems, that performed the reduction of the angular momentum, are singular when the eccentricities tend to zero.
} \\
$(\theta_j, J_j, x_j, -i\xb_j, y_j, -i\yb_j)$ with  
\be
\begin{disarray} {ll}
\theta_1 = \lam_1 - \lam_2,   & 2J_1  = \Lam_1 - \Lam_2,\quad \\
\theta_2 = \lam_1 + \lam_2,  & 2J_2 = \Lam_1 + \Lam_2.
 \end{disarray}
 \ee

Inside the 1:1 mean motion resonance, the angular variable $\theta_1$ varies slowly with respect to $\theta_2$. Consequently,  the planetary Hamiltonian (\ref{eq:ham_cart}) will be averaged over the angle $\theta_2$.   
\subsection{The quasi-circular and planar average problem}
\subsubsection{The average problem}
\label{sec:average}
 
In this paper, we only consider  the average Hamiltonian at first order in the planetary masses. More precisely, we assume that there exists a canonical transformation which maps the initial Hamiltonian $H$ to 
\be
\Hb(\theta_j,J_j, x_j, -i\xb_j, y_j, -i\yb_j) =  \Hb_0(J_j)  + \Hb_1(\theta_1, J_j, x_j, -i\xb_j, y_j, -i\yb_j)   + \gO(\eps^3) 
\ee
with
\be
\Hb_0(J_1,J_2) = -\frac{\beta_1^3\mu_1^2}{2(J_1+J_2)^2} -\frac{\beta_2^3\mu_2^2}{2(J_1-J_2)^2} 
= H_K\circ\phi(\theta_j, J_j, x_j, -i\xb_j, y_j, -i\yb_j) 
\ee
and
\be
\Hb_1(\theta_1, J_j, x_j, -i\xb_j, y_j, -i\yb_j) = \frac{1}{2\pi}\int_0^{2\pi} H_p\circ\phi(\theta_j, J_j, x_j, -i\xb_j, y_j, -i\yb_j) d\theta_2 \, ,
\label{eq:Hbar}
\ee
where the map $\phi$ satisfy the relation $(\brt_j,\br_j) =\phi(\theta_j, J_j, x_j, -i\xb_j, y_j, -i\yb_j)$. If we denote by  $(\theta_j, J_j, x_j, -i\xb_j, y_j, \\ -i\yb_j)$ the canonical variables associated to the average problem, we remark that $J_2 =  \Lam_1 + \Lam_2$ is a first integral of $\Hb$. It is also easy to prove that the quantities $\sum_j y_j\sqrt{\Lam_j  - \vert x_j\vert^2 - \vert y_j \vert^2/2  }$ and $\sum_j \left[ \Lam_j  - \vert x_j\vert^2 - \vert y_j \vert^2\right]$ are first integrals too.  It is possible to take advantage of these first integrals by reducing the problem by means of adapted canonical coordinate system as it is the case with the Jacobi reduction in the spatial problem (see \cite{Ro1995} and \cite{MaRoLa2002}).   The reduction can also be achieved in the planar problem leading to two degrees of freedom Hamiltonian system depending on two angles:  the difference of the mean longitudes and the difference of the longitudes of the perihelion (\cite{GiuBeMiFe2010}).  These reductions introducing some technical issues (addition of a parameter, singularities when the eccentricities and inclinations tend to zero), we prefer not to reduce the problem.

The average Hamiltonian (\ref{eq:Hbar}) depending on the mean longitude only by their difference $\theta_1$, the rotational invariance of the Hamiltonian given by the relation (\ref{D'Al})  imposes that $\Hb$ is even in the variables $x_j$ and $y_j$ and their conjugates. 
As a consequence, the set $x_1 = x_2 = y_1 = y_2 =0 $ is an invariant manifold by the flow of  the average Hamiltonian (\ref{eq:Hbar}).  More generally, this property holds for any order of averaging. 
This implies that the part of the average Hamiltonian (\ref{eq:Hbar}) which does not depend on  the eccentricities and the inclinations, namely  $H_0(\theta_1, J_j) =  \Hb(\theta_1, J_j, 0,0,0,0)$, is an integrable Hamiltonian. 
It is worth noting that the one degree of freedom Hamiltonian $H_0$, associated to the circular and planar resonant problem, is a peculiar attribute of the 1:1 mean-motion resonance.  The next section is devoted to its study.

\subsubsection{The integrable part $H_0$}
\label{sec:partH0}
After replacing the vectors $\br_j$ and $\brt_j$ by their expressions  in terms of elliptic elements into the planetary Hamiltonian (\ref{eq:ham_cart}),  
an explicit expression of $H_0$ is obtained by suppressing the terms depending on the variables $x_j, \xb_j, y_j, \yb_j$ and the fast angle $\theta_2$.   
This leads to the Hamiltonian
\be
\begin{split}
H_0 = &-\frac{\beta_1\mu_1}{2a_1} - \frac{\beta_2\mu_2}{2a_2} \\
&+ \cG m_1m_2\left(
\frac{\cos\theta_1}{\sqrt{a_1a_2} } -\frac{1}{\sqrt{a_1^2+a_2^2 -2a_1a_2\cos\theta_1}}
\right),
\end{split}
\label{eq:H_0planetes}
\ee
where the semi-major axis $a_j$ depends on the action $J_1$ and the first integral $J_2$.
The constant  $\tpr{2}J_2 = \Lam_1 + \Lam_2$ being positive, there exists a strictly positive number  $\ab$ such that 
\be
J_2 = \frac{ \beta_1\sqrt{\mu_1} + \beta_2\sqrt{\mu_2} }{2} \sqrt\ab.
\ee
At this point, it is convenient to define a new couple of conjugate variables $(\theta, J)$ by translating the action $J_1$ as
\be
J_1 = \frac{ \beta_1\sqrt{\mu_1} - \beta_2\sqrt{\mu_2} }{2} \sqrt\ab + J, \quad \theta_1 = \theta  \,.
\ee
It will also be  useful to define the dimensionless (non canonical)  action-like variable $u$  by the relation
\be
J = (\beta_1+\beta_2)\sqrt{\mu_0\ab} \, u \qtext{with} \mu_0 = \cG m_0.
\ee
Now, by a substitution of the relations
\be 
 a_j = \left( \sab  +  \frac{(-1)^{j+1}}{\beta_j\sqrt{\mu_j}} J\right)^2 = 
 \ab\left( 1  + (-1)^{j+1} \frac{\beta_1+\beta_2}{\beta_j}\sqrt{\frac{\mu_0}{\mu_j}  } u\right)^2
 \label{eq:aJu}
 \ee 
into the expression (\ref{eq:H_0planetes}),  the integrable average Hamiltonian $H_0$ can be explicitly expressed in terms of the $(\theta,J,\ab)$, or $(\theta, u,\ab)$ for convenience. 
Note that the expression (\ref{eq:aJu}) allows one to interpret the parameter $\ab$ as a mean value around which the semi-major axes oscillate. 

\begin{figure}[h!]
\includegraphics[width=12.5cm]{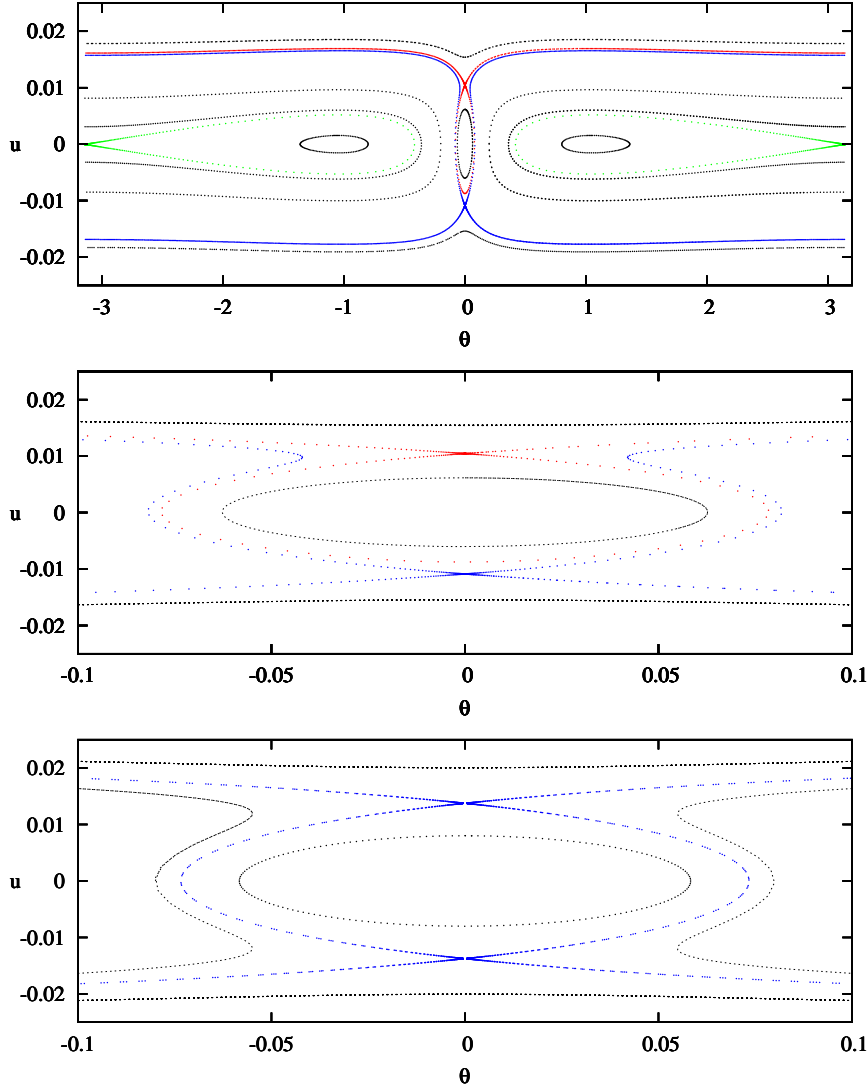}
\caption{Phase portrait of the Hamiltonian $H_0$ in coordinates $(\theta, u)$. The upper box shows the whole space for $m_0 =1$,   $m_1 = m_J$ and $m_2 = m_S$, the parameters $\cG$ and $\ab$ being equal to one. The separatrix that originates at $L_3$ ($\cS_3$) is plotted in green, while $\cS_2$  is the blue curve and $\cS_1$ the red one. The middle plot is an enlargement of the region surrounding the collision point while the bottom plot shows the merging of the two separatrices  $\cS_1$ and $\cS_2$ when the planetary masses are equal. Here, their values are $m_1=m_2 = 5\times 10^{-4}$.
}
\label{fig:figH0}
\end{figure}

The figure \ref{fig:figH0} reproduces the phase portrait of the integrable Hamiltonian $H_0$ in coordinates $(\theta, u)$.  It can easily  be expressed in terms of semi-major axes using the expression (\ref{eq:aJu}) or their first order approximation $a_j-\ab \approx \left[2(-1)^{j+1}\ab (m_1+m_2)/m_j \right] u$.   
The upper plot represents the whole phase diagram for $m_1 = m_J=10^{-3}$ and $ m_2 = m_S = 3\times10^{-4}$ and $\cG = m_0 = \ab =1$, where the masses $m_J$ and $m_S$ are close to those of Jupiter and Saturn expressed in solar mass. This plot is similar to the well known Hill's diagram (or zero-velocity curves) of the non averaged planar circular RTBP (see \cite{Sze1967}) although the zero-velocity curves are not solution curves of the motion. It is also topologically equivalent to the phase space of the average planar circular RTBP when the eccentricity of the test-particle is equal to zero \citep{NeThoFeMo02, Morbidelli02}.
The Hamiltonian system associated to $H_0$ possesses five fixed points that correspond to the usual Euler and Lagrangian configurations, and one singular point at $u = \theta = 0$ which corresponds to the collision between the planets.  
The two stable equilibrium points located at $\theta = \pm\pi/3, u = 0$ (see the next paragraph for  more details) represent the average equilateral configurations that we will denote abusively by $L_4$ and $L_5$ by analogy with the RTBP. Each of these points is surrounded by tadpole orbits corresponding to periodic deformations of the equilateral triangle. This region is bounded by the separatrix $\cS_3$ that originates at the hyperbolic fixed point $L_3$ at $\theta = \pi, u \approx 0$, for which the three bodies are aligned and the Sun is between the two planets and its separatrix. Outside of this domain, the horseshoe orbits are enclosed by the separatrix $\cS_2$  that originates at the fixed point $L_2$  ($\theta =0$ and $u < 0$). This point, as the equilibrium point  $L_1$, is associated with an Euler configuration for which the two planets are on the same side of the Sun.  The last domain, centered at the singularity, is surrounded by the  separatrix $\cS_1$ connecting the $L_1$ point ($\theta =0$ and $u>0$) to itself. Inside this small region,  the two planets seem to be subjected to a prograde satellite-like motion, the one revolving the other one clockwise. 
By an enlargement of this region ( $-0.1<\theta < 0.1$), the second plot of Fig. \ref{fig:figH0} (middle box) shows the splitting of the two separatrices $\cS_1$ (red) and $\cS_2$  (blue) when the planetary masses are different.
On the contrary, for equal planetary masses, the phase portrait becomes symmetric with respect to the axis $u = 0$. It turns out that the equilibrium points $L_3, L_4, L_5$ lie on the axis of symmetry, and that the two curves $\cS_1$ and $\cS_2$ merge together giving rise to a unique separatrix connecting $L_1$ to $L_2$. The bottom plot of Fig. \ref{fig:figH0} describes this phenomenon for $m_1 = m_2 = 5\times10^{-4}$, the other parameters being unchanged.  

A way to  estimate  the locations of the equilibrium points is to use an asymptotic expansion of the Hamiltonian (\ref{eq:H_0planetes}) in the neighborhood of $u=0$. Two cases have to be considered. The first one arises in a domain which excludes a suitable neighborhood of the collision  (the distance between the planets has to be of order unity).   The second case concerns a small domain enclosing the singularity. 
In the first situation, our goal can be achieved by an expansion  of  the Hamiltonian (\ref{eq:H_0planetes}) in the neighborhood $u=0$, assuming that the condition  $\theta = \gO(1)$ is fulfilled. We will see later that this is satisfied in the tadpole region. This condition   also holds for the horseshoe orbits which do not approach too much the singularity. 
 Denoting by $\Gam$ the quantity $ \sqrt{2-2\cos\theta}$, and using the notations
 $\sigma_1 = m_1+m_2$, $\sigma'_1 = m_1 - m_2$ and $\sigma_2 = m_1m_2$, the expansion of $H_0$ can be written as
\be
\begin{split}
H_0(\theta,u) = \cG\ab^{-1}
&\left[
\gam_1 + \gam_2(\theta) +  \gO(\eps^3) 
+\left( \delta_2(\theta) +  \gO(\eps^3) \right) u 
\right.  \\
 &\left.
 + \left(\tau_1 + \tau_2(\theta) +   \gO(\eps^3) \right) u^2
+ u^3R(\theta,u,\eps)
\right]\,,
\end{split}
\label{eq:dlHamReg}
\ee
where the coefficients $\gam_k, \tau_k, \delta_k$ are given by
%
%
%
\be
\begin{split}
&2\gam_1 = -m_0\sigma_1,  \quad
 2\gam_2 = \sig_2(2 -\Gam^2-2\Gam^{-1}),\\
&2\delta_2  =  - \sig_1\sig'_1(1-\Gam)^2(1+2\Gam^{-1}), \quad
2\tau_1 = -3 m_0\sig_1^3\sig_2^{-1},\\ 
&\tau_2 = \sig_1^2\sig_2^{-1}
\left[
(\sig_1^2-3\sig_2)(4 -\Gam^2/2 -\Gam^{-1}) + 2\sig_1^2\Gam^{-3}
\right], 
\end{split}
\label{eq:dlHamRegCoef}
\ee
 and the remainder $R$ is a periodic function of $\theta$ depending on $u$ and of order $\eps$.
The location of the fixed points $L_3$, $L_4$ and $L_5$, as well as the eigenvalues of the associated linearized system, can be easily deduced from the expansion (\ref{eq:dlHamReg}). 
The location of the two elliptic fixed points $L_4$ and $L_5$ is approximated by  $(\theta,u) = (\pm \pi/3, 0 + \gO(\eps^2))$, which leads to $a_1 = \ab( 1 + \gO(\eps^2)),  a_2 = \ab(1 + \gO(\eps^2))$. The quadratic expansion of the Hamiltonian $H_0$ in the neighborhood of $L_4$ (change $\pi/3$ in $-\pi/3$ for $L_5$)  is equal to
\be
H_{0, L_4}^{(2)} = 
-\frac32\frac{\cG}{\ab}\left(
 \frac{\sig_1^2}{\sig_2}(m_0\sig_1 - 3\sig_1^2 +5\sig_2)u^2 + \frac34\sigma_2 \left(\theta -\frac{\pi}{3}\right)^2
\right),
\label{eq:dlLag}
\ee
where only the dominating terms in $\eps$ are retained. Moreover, the frequency associated to this elliptic fixed point  reads
%
\be
\nu_0 = n_0\sqrt{\frac{27}{4} \frac{\sig_1}{m_0}}\left( 
         1 - \frac{\sig_1^2-\sig_2}{2m_0\sig_1} + \gO(\eps^2)
          \right),
\ee
where $n_0 = \mu_0^{1/2}\ab^{-3/2}$ plays the role of an averaged  mean motion. 
The location of the hyperbolic point $L_3$ is  obtained by translating  $u$ by a quantity $u^{(3)}$ which cancels the  linear term in  $u$ in  the expansion (\ref{eq:dlHamReg}) when $\theta = \pi$. We get the approximation
\be
u^{(3)} =  -\frac{(m_1-m_2)m_1m_2}{3m_0(m_1+m_2)^2} + \gO(\eps^2),
\label{eq:u3}
\ee
 which gives in terms of semi-major axes:
\be
a_j = \tpr{\ab} \left(
1 + \frac{(-1)^j}{3}\frac{m_k}{m_0}\frac{m_1-m_2}{m_1+m_2} + \gO(\eps^2)
\right) \qtext{with }Êj\neq k. 
\label{eq:a_u}
\ee
As a consequence, the quadratic expansion of $H_0$ in the vicinity of $L_3$, whose coordinates are $(\pi,u^{(3)})$, reads
\be
H_{0, L_3}^{(2)} =
 -\frac32\frac{\cG}{\ab}\left(
 \frac{\sig_1^2}{\sig_2}(m_0\sig_1 - \frac76\sig_1^2 +3\sig_2)(u-u^{(3)})^2 - \frac{7}{24}\sigma_2 \left(\theta - \pi \right)^2
\right).
\label{eq:dlEul}
\ee

The domain including the tadpole orbits is bounded by the separatrix  $\cS_3$. The size of this domain can be estimated by different ways. A simple manner to achieve this goal is to calculate the quantity $U_3$ which is the maximal value taken by the action $u$ along $\cS_3$. By solving the equation $H_0(\pi/3,U_3) = H_0(\pi, u^{(3)})$ where $H_0$ is approximated by  (\ref{eq:dlHamReg}), we get the expression:
\be
U_3 = \frac{\sqrt 2\sig_2}{\sqrt{3 m_0\sig_1^3}} + \gO(\eps^2) = \frac{\sqrt2 m_1 m_2}{\sqrt{3m_0(m_1+m_2)^3}} +\gO(\eps^2).
 \label{eq:uL3}
\ee
Similarly,  in the perpendicular direction, we can also estimate the quantity $\Theta_3$ corresponding to the minimal value of $\theta$ along $\cS_3$ by an approximation of the positive root of the equation $H_0(\Theta_3,0) = H_0(\pi, u^{(3)})$. The solution is given by
\be
\Theta_3 = 2\arcsin(\frac{\sqrt{2}-1}{2}) + \gO(\eps) \approx 23.9^{\circ},
\ee
which is a classical result in the case of the RTBP \citep{Ga1977}.
 
The approximation (\ref{eq:dlHamReg}) is not valid for the Euler points $L_1$ end $L_2$, the latter being located at a distance of order $\eps^{1/3}$ of the singularity. 
In this case, we can use the  asymptotic expansion of $H_0(0,u)$, valid since $u = \gO(\eps^\alpha)$ with $0 \leq \alpha <1$, given by 
\be
\begin{split}
H_0(0,u) = \cG\ab^{-1}
&\left[
\gam'_{-1}{\vert u\vert}^{-1} + \gam'_1 + \gam'_2 +  \gO(\eps^3) 
+\left( \delta'_2 +  \gO(\eps^3) \right) u 
\right.  \\
 &\left.
 + \left(\tau'_1 + \tau'_2 +   \gO(\eps^3) \right) u^2
+ u^3R'(u,\eps)
\right],
\end{split}
\label{eq:dlHamSing}
\ee
where the coefficients $\gam_k, \tau_k, \delta_k$ are
\be
\begin{split}
&2\gam'_{-1} = -\sig_1^{-2}\sig_2^2, \quad 2\gam'_1 = 2\gam_1 = -m_0\sigma_1,  \quad
 \gam'_2 = \sig_2, \\
&2\delta_2  =  3\sig_1\sig'_1, \quad
2\tau_1 = 2\tau'_1 = -3 m_0\sig_1^3\sig_2^{-1}, \\ 
&\tau'_2 =  4(\sig_1^2-3\sig_2)\sig_1^2\sig_2^{-1}
\end{split}
\label{eq:dlHamSingCoef}
\ee
and $R'$ is a periodic function of $\theta$ depending on $u$ and of order $\eps$.
At this  accuracy, the two Euler points $L_1$ and $L_2$ are symmetric with respect to the line $u = 0$, and if we denote by  $u^{(1)}$ (resp. $u^{(2)}$)  the $u$-coordinate of $L_1$ (resp. $L_2$),  we have $u^{(2)} = - u^{(1)} + \gO(\eps^{2/3})$ and
\be
u^{(1)} = \frac{\sig_2 }{(6m_0\sig_1^5)^{1/3}}+ \gO(\eps^{2/3}) = \frac{m_1m_2}{(6m_0(m_1+m_2)^5)^{1/3}} + \gO(\eps^{2/3})
\label{eq:u1}.
\ee
Let us mention that the quantities $u^{(1)}, u^{(2)}$ and $u^{(3)}$ can also be considered as roots of a polynomial equation, as it is the case for  Euler's configurations in the full three-body problem (see  \cite{MaBo1982} or \cite{Roy1982}).

 As for the tadpole orbits, the  width of the horseshoe region along the $u$ axis can be deduced from the equation  $H_0(\pi/3,U_1) = H_0(0, u^{(1)})$ that is 
 \be
 U_1 = 2^{-1/2}6^{1/6} m_0^{-1/3}\sig_1^{-5/3}\sig_2 + \gO(\eps^{2/3}).
 \label{eq:uL1}
  \ee
 The minimal angular separation between two planets in horseshoe orbit is solution of the equation $H_0(\theta^{(1)},0) = H_0(0, u^{(1)})$, and is equivalent to
 \be
 \theta^{(1)} = \frac43\left(\frac{\sig_1}{6m_0}\right)^{1/3} + \gO(\eps^{2/3}).
 \ee
 We note that at this degree of accuracy (neglecting the terms of order $\eps^{2/3}$), the two separatrices $\cS_1$ and $\cS_2$ merged in a single curve. 

The  equations (\ref{eq:uL3}) and (\ref{eq:uL1}) allow one to retrieve the result on the relative size of the tadpole and horseshoe regions obtained by \cite{DeMu1981a}  in a very different way.   Indeed, we have
\be
\frac{U_3}{U_1} = 2\sqrt[6]{6}\left(\frac{m_1+m_2}{m_0} \right)^{1/6} = \gO(\eps^{1/6}).
\ee
As a consequence, the lower the planetary masses are,  the larger the horseshoe region is (with respect to the width of the tadpole region). 

It is worth to mention that,  although the average system modeled by the Hamiltonian $H_0$ provides a faithful representation of the topology of the problem, it only reflects poorly the dynamics in the  domain  bounded by $\cS_1$ containing the singularity.

The simplest argument that points out this problem comes from the computation of the  orbit frequency surrounding the collision point. Indeed, close to the singularity, the Hamiltonian can be approximated by
\be
-\cG m_1m_2\ab^{-1}(\alpha J^2 + \theta^2)^{-1/2} \qtext{with} \alpha = (m_1m_2\mu_0\ab)^{-1}.
\ee
It turns out that the frequency of the trajectory that originates at $\theta =\theta_0$ and $J = 0$ is equivalent to 
\be
\frac{\sqrt{m_1m_2}}{m_0} \frac{n_0}{\theta_0^3}.
\ee
As a  product of an averaging process, this frequency would be small compared to $n_0$, but it tends to infinity when $\theta_0$ tends to zero.

\begin{figure}[h!]
\begin{center}
\includegraphics[width=5.5cm,height=5.6cm]{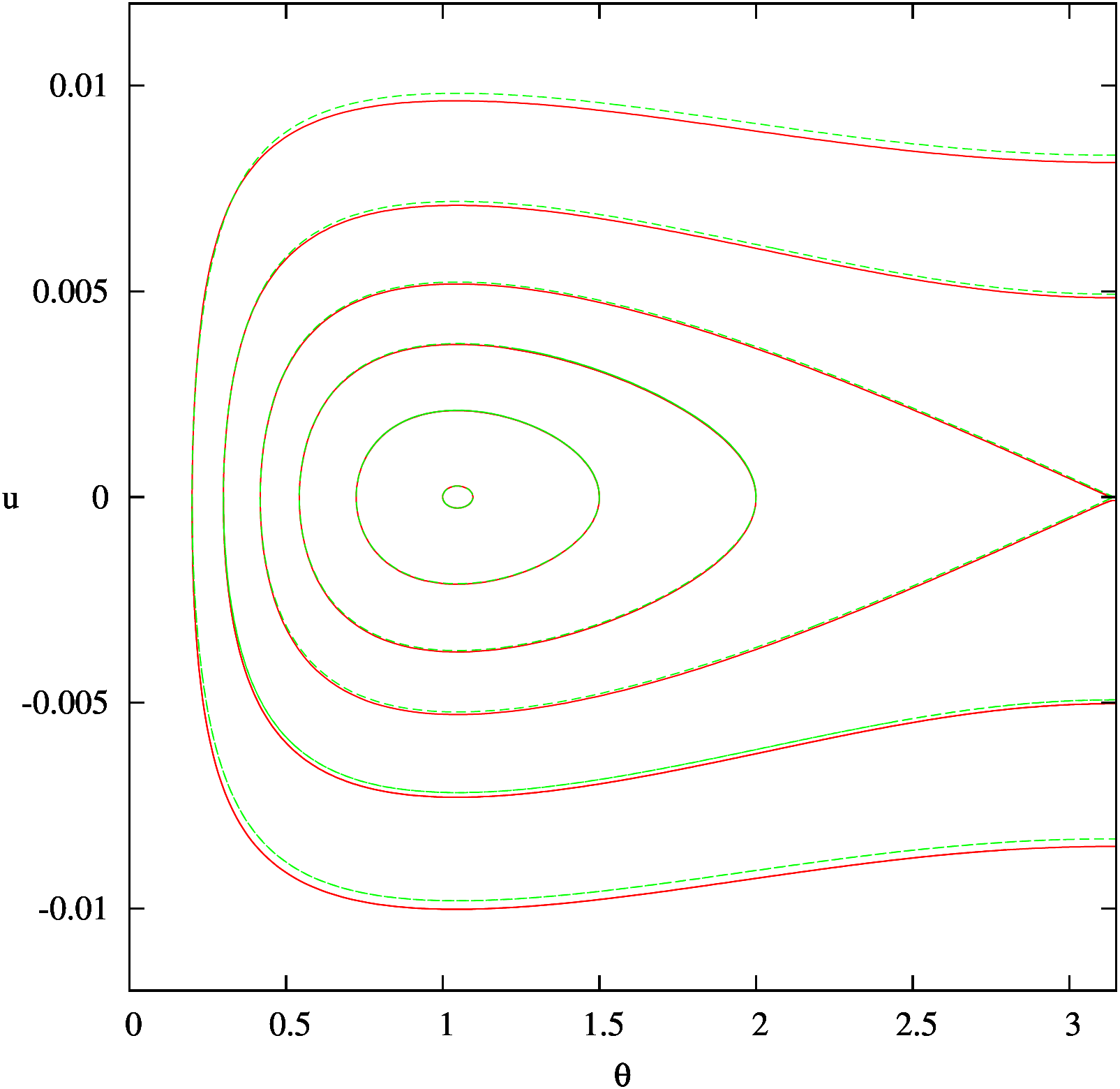} 
\hspace{15pt}
\includegraphics[width=5.5cm,height=5.6cm]{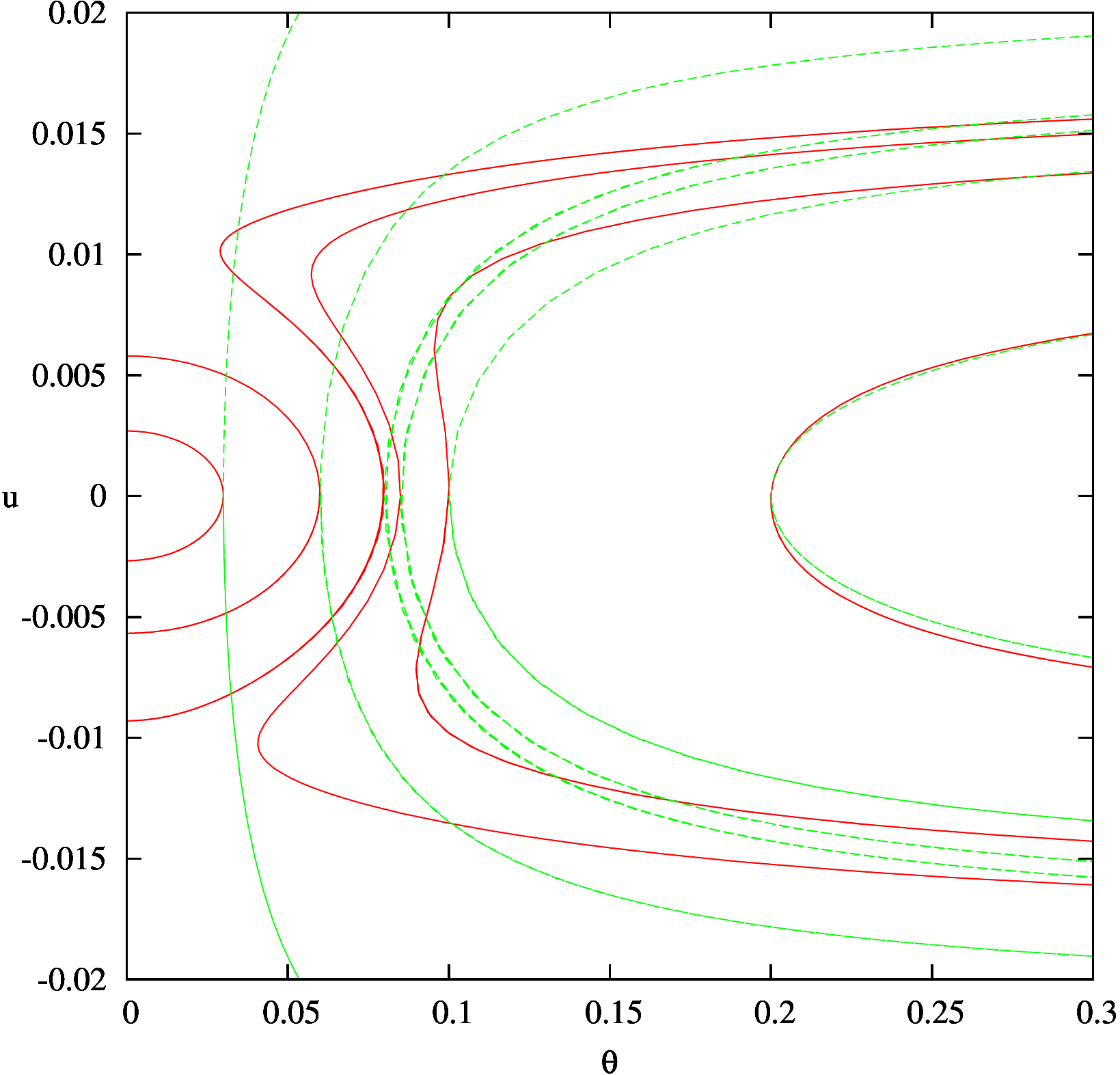} 
\end{center}
\caption{Comparison between the model $H_0$ and its approximation $H_a$. The level curves of $H_0$ are plotted in red while the level curves of $H_a$ are in green. The two approximations fit correctly when the trajectories do not come too close to the singularity (left panel). The right plot shows the inconsistency between the two models in the Hill region. }
\label{fig:figCompar}
\end{figure}

To conclude this section, we will  compare the average Hamiltonian $H_0$ to  a classical approximation of the co-orbital resonance. This model, represented by the Hamiltonian $H_{a}$ which reads
\be
H_{a} =  -\frac32\frac{\cG}{\ab}\tpr{\frac{m_0(m_1+m_2)^3}{m_1m_2}}u^2 + \frac{\cG m_1m_2}{\ab}\left(\cos\theta - \frac{1}{\sqrt{2-2\cos\theta}}\right),
\label{eq:Happrox}
\ee
 has been used by  \cite{YoCoSyYo1983} to study the dynamics of the co-orbital satellites Janus and Epimetheus \tpr{(see also \cite{SiDu2003})  }.  A similar Hamiltonian is also employed to model the 1:1 mean motion resonance in the RTBP (see \cite{Morais1999}).
This approximation is a particular case of the expression (\ref{eq:dlHamReg}) obtained by expanding  $H_0$ in power series of $u$ and $\eps$. For small enough values of $u$, and if $\theta$ is not too close to $0$ or to  $2\pi$, the Hamiltonian $H_a$ provides a good approximation of $H_0$ in the tadpole region and  for horseshoe orbits providing $u<<\eps^{1/3}$.  
As one can see on the left plot of the  figure \ref{fig:figCompar}, although the differential system associated to the Hamiltonian $H_a$ possesses only three fixed points, its trajectories are very close to those of the average Hamiltonian $H_0$, when they do not approach the collision. This is especially true for the tadpole orbits and the moderate amplitude horseshoe orbits. 
On the contrary, as shown in Fig. \ref{fig:figCompar} (right panel), the second model is not valid in the Hill region. But we have to keep in mind that though in this region the topology of the average problem is well described by $H_0$, it is not the case of its dynamics.

\section{Variational equations in the neighborhood of the quasi-circular problem}
\label{sec:quadratic}

\subsection{The variational equations}
\label{sec:fixedpoints}
It has been shown in the previous section that the manifold $ x_j= y_j = 0$ is invariant by the flow of the  average Hamiltonian (\ref{eq:Hbar}). In order to study the (linear) stability of this invariant manifold in the transversal directions $(x_j, y_j)$, we have to calculate  the variational equations associated to this invariant surface.  These equations, corresponding to the linearization of the differential system associated to the Hamiltonian (\ref{eq:Hbar}) in the neighborhood of the plane $ x_j= y_j = 0$, can be derived from the quadratic expansion  in eccentricity and inclination of the average Hamiltonian $\Hb$. This expansion can be written in the form  $H_0 + H_2^{(h)} + H_2^{(v)}$ with 
\be
H_2^{(h)} =  \cG m_1m_2\left( A_h X_1\Xb_1 + B_hX_1\Xb_2 + \Bb_h\Xb_1X_2 + A_hX_2\Xb_2 \right)
\label{eq:Hquadh}
\ee
and 
\be
H_2^{(v)} =  \cG m_1m_2\left(  A_vY_1\Yb_1 + B_vY_1\Yb_2 + \Bb_v\Yb_1Y_2 + A_vY_2\Yb_2 \right),
\label{eq:Hquadv}
\ee
where the coefficients $A_h, B_h, A_v$ and $B_v$ read
\be
\begin{split}
A_v &= \left(\frac{a_1a_2}{\Delta^3} - \frac{1}{\sqrt{a_1a_2}}\right)\cos\theta,\quad
B_v = \left(\frac{1}{\sqrt{a_1a_2}} - \frac{a_1a_2}{\Delta^3} \right)e^{i\theta}, \\
A_h &= \frac{a_1a_2}{8\Delta^5}\left(
a_1a_2(5\cos2\theta - 13) + 4(a_1^2+a_2^2)\cos\theta
\right)
 - \frac{\cos\theta}{2\sqrt{a_1a_2}}, \\
 B_h &=  \frac{e^{-2i\theta}}{2\sqrt{a_1a_2}}  -\frac{a_1a_2}{16\Delta^5}
 \left(
 a_1a_2\left( e^{-3i\theta} + 9 e^{i\theta} -26e^{-i\theta}  \right)+ 8(a_1^2+a_2^2)e^{-2i\theta}
 \right), \\
 \Delta & = \sqrt{a_1^2+a_2^2 -2a_1a_2\cos\theta}.
 \end{split}
 \label{eq:coeffHquad}
\ee
The formulas (\ref{eq:Hquadh}),   (\ref{eq:Hquadv}) and (\ref{eq:coeffHquad}) generalize the expansion given by  \cite{Morais1999,Morais2001} in the case of the elliptic RTBP. 

The variational equations in the vicinity of a solution lying in the plane $x_j = y_j =0$ and satisfying
\be
\dot \theta = \frac{1}{c}\dron{H_0}{u}(\theta,u), \quad \dot u =  -\frac{1}{c}\dron{H_0}{\theta}(\theta,u) \qtext{with} c = (\beta_1+ \beta_2)\sqrt{\mu_0\ab}
\label{eq:systutheta},
\ee
take the form
\be
\vect{\dot X_1}{\dot X_2}{} = 2i\cG m_1m_2
\left(\begin{array}{ll}
\Lam_1^{-1}A_h   & \Lam_1^{-1}\Bb_h \\
\Lam_2^{-1}B_h   & \Lam_2^{-1}A_h
\end{array}\right)
\vect{ X_1}{ X_2}{}
=
M_h(\theta,u) \vect{ X_1}{ X_2}{}
\label{eq:variationH}
\ee
and
\be
\vect{\dot Y_1}{\dot Y_2}{} = \frac{i\cG m_1m_2}{2}
\left(\begin{array}{ll}
\Lam_1^{-1}A_v   & \Lam_1^{-1}\Bb_v \\
\Lam_2^{-1}B_v   & \Lam_2^{-1}A_v
\end{array}\right)
\vect{ Y_1}{ Y_2}{}
=
M_v(\theta,u) \vect{ Y_1}{ Y_2}{},
\label{eq:variationV}
\ee
where $\theta$ and $u$ are deduced from the solutions of the equations (\ref{eq:systutheta}), and the $\Lam_j$ implicitly depend on $u$ by the relations (\ref{eq:aJu}).
As these solutions  are periodic (except if their initial conditions are chosen on the separatrices $\cS_1$ to $\cS_3$) the linear equations (\ref{eq:variationH}) and (\ref{eq:variationV})  are periodically time-dependent. As a consequence, their solutions cannot  generally  be expressed in a close form. A notable exception occurs at the equilibrium points of the system  (\ref{eq:systutheta}). Indeed, here, the variational equations become autonomous and consequently integrable. Then we will first begin to study  these special cases. 
Before going further, let us mention that in this paper, we will not study the ``vertical" variational equation (\ref{eq:variationV}).
Indeed, due to its strong degeneracy, the study of this linear equation is not sufficient to understand the local dynamics. To this aim, the use of higher order terms of the Hamiltonian is necessary (at least the forth degree in $y_j$).  To be convinced, it is enough to look at the matrix $M_v$, defined in (\ref{eq:variationV}),  for $\theta = \pi/3$ and $u = 0$. Indeed, at $L_4$ the matrix vanishes and the quadratic Hamiltonian does not provide any information about the dynamics in the $y_j$ directions.
Then, this situation requires a careful analysis of the structure of the Hamiltonian in order to deal with potential bifurcations in the vertical direction, as it is pointed out by  \cite{Jo2000} in the case of the bicircular problem. 
Therefore, we postpone this study to a future work. 

\subsubsection{Dynamics around the fixed points}
\label{sec:L4lin}
For the equilateral configurations ($\theta = \pm \pi/3$),  neglecting the quadratic terms  in $\eps$, the matrix $M_h$ takes the following expression
\be
 M_h =  -i\frac{27}{8}\dfrac{n_0}{m_0}
 \left(\begin{array}{cc}
m_2  & -m_2e^{i\theta} \\
-m_1e^{-i\theta}   & m_1
\end{array}\right).
\ee
This matrix possesses two eigendirections associated to the eigenvectors 
\be
V_1 = \vect{ e^{i\theta} m_2}{ -m_1}{} \qtext{and}
V_2 = \vect{ e^{i\theta} }{ 1}{},
\label{eq:eigenVectEqui}
\ee
whose eigenvalues are respectively 
\be
v_1 = -i\frac{27}{8}\frac{m_1+m_2}{m_0} n_0 \qtext{and}  v_2 = 0.
\ee
These eigenvectors have a precise physical meaning. Along the neutral direction, the one which is collinear with $V_2$, the two eccentricities are the same and the angle $\Delta\varpi = \varpi_1 - \varpi_2$ separating the two apsidal lines is equal to $\pi/3$ at $L_4$ and $-\pi/3$ at $L_5$. These configurations clearly correspond to the Lagrangian elliptic equilibria, which are fixed points of the average problem, and consequently of the linearized average problem at $L_4$ or $L_5$. This is the reason why the associated eigenvalue $v_2$ vanishes. 
Along the direction $V_1$, the orbits satisfy the relations 
\be
a_1 = a_2 = \ab,\quad \theta = \pm\pi/3, \quad
 m_1e_1 = m_2e_2,   \qtext{and}  \Delta\varpi = \varpi_1 - \varpi_2 = \theta + \pi.
\ee
This corresponds to an infinitesimal version of the Anti-Lagrange orbits found numerically by  \cite{GiuBeMiFe2010}.
On these trajectories the elliptic elements $a_1, a_2,e_1, e_2$ and $\theta$ are constant. Only the two angles $\varpi_1$ and $\varpi_2$ precess with the same frequency equal to \be
g_1 = iv_1 = \frac{27}{8}\frac{m_1+m_2}{m_0} n_0,
\ee
in such a way that the angle $\Delta\varpi$ is constant. 
As a consequence, this family of periodic orbits is transformed, after reduction by the rotations, in a family of fixed points, which is exactly what have found \cite{GiuBeMiFe2010} in the reduced problem. Of course, the family that we found along the eigenvector $V_1$ of the linearized system provides only an infinitesimal approximation of the Anti-Lagrange family in the neighborhood of $L_4$,  but we will show in Section \ref{sec:LypFam} that this linear approximation can be generalized to any degree using Birkhoff normal form.

 \subsubsection{The Eulerian fixed point $L_3$ }  
 By evaluating the matrix $M_h(\theta,u)$ at   $(\theta,u) = (\pi,u^{(3)})$ and neglecting the terms in $\eps^2$ and more, the matrix of the linearized system at $L_3$ reads
\be
 M'_h = i\frac{7}{8}\dfrac{n_0}{m_0}
 \left(\begin{array}{cc}
m_2  & m_2 \\
m_1   & m_1
\end{array}\right).
\ee
This matrix possesses two eigendirections associated to the eigenvectors 
\be
V'_1 = \vect{m_2}{ m_1}{} \qtext{and}
V'_2 = \vect{1}{ -1}{},
\label{eq:eigenVectEuler}
\ee
whose eigenvalues are respectively 
\be
v'_1 =  i\frac{7}{8}\frac{m_1+m_2}{m_0} n_0 \qtext{and}  v'_2 = 0.
\ee
 As in the equilateral case, the direction $V_2'$ corresponds to the unstable Euler configurations where the two planets are in the two sides of the Sun (the eccentricities are equal and the perihelia are in opposition). 
 The other direction is more interesting. In this case, the perihelia are in conjunction and the eccentricities verify the relation $m_1e_1 = m_2e_2$.   As for $L_4$,   the method developed  in Section 4.1 makes possible to prove the existence of  a one-parameter family of periodic orbits that bifurcates from $L_3$  and is tangent to $V'_1$ at this point.  Remark that, at least close to $L_3$, this family has been numerically  computed  by \cite{HaPsyVo2009}, and were previously found  by \cite{Poi1892} as a  solution of second sort \footnote{M\'ethodes nouvelles de la m\'ecanique celeste Vol I, Chap III, \&47:``Solutions de la seconde sorte".} (see \cite{chen2012}).

\subsubsection{Euler $L_1$ and $L_2$ equilibria} 
As the Hamiltonian $H_0$ does not reflect properly the dynamics in the neighborhood of the collision between the two planets, the linearized problem at $L_1$ or $L_2$ will not be considered in the present paper (see the end of Section \ref{sec:partH0}). 

\subsection{The general solution of the variational equation}

\begin{figure}[h!]
\includegraphics[width=12cm]{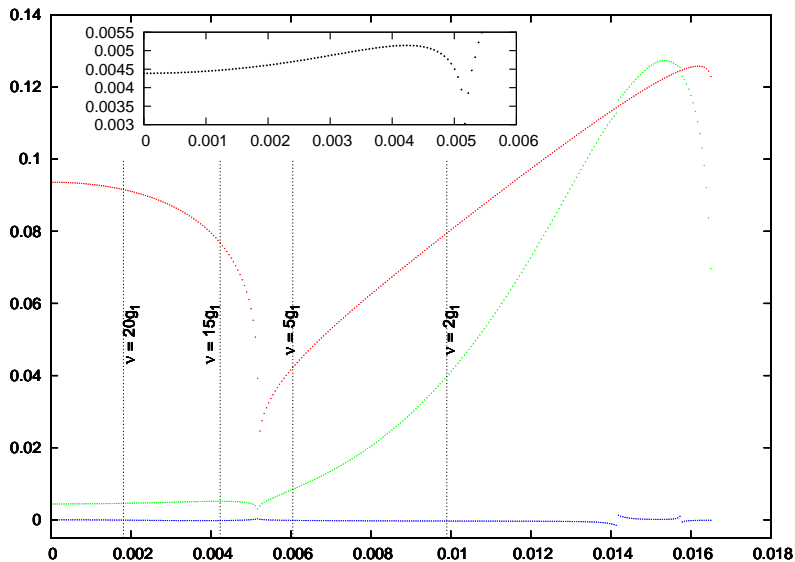} 
\caption{Variations of the fundamental frequencies of the variational equation.  $u$ is plotted along the X-axis and the frequencies, in rad/yr, along the Y-axis. The red curve represents the evolution of the libration frequency $\nu$ along the segment $\theta = \pi/3$, $0\leq u < u^{(1)}$, while the green (resp. blue) curve is associated to $g_1$ (resp. $g_2$). The vertical black dotted lines indicate the location on some resonances between $g_1$ and $\nu$. The box plotted in the upper left corner of the figure is an enlargement showing the behavior of $g_1$ for $0\leq u < 0.0055$. }
\label{fig:freqH0}
\end{figure}

Now, let us study the general case. This corresponds to writing the variational equation (\ref{eq:variationH}) around a periodic solution of frequency $\nu$.  According to the Floquet theorem (see  \cite{MeHa1992}), the solutions of the variational equation take the form
\be
z(t) = P(\nu t) \exp(At),
\ee
where $A$ is a constant matrix and $P(\psi)$ is a matrix whose coefficients are $2\pi$-periodic functions of $\psi$. 
As, if $Z$ is a fundamental matrix solution to the variational equation along a $2\pi/\nu$-periodic solution,  one has the relation 
\be
Z(t + 2\pi\nu^{-1}) = Z(t)\exp\left(
2\pi\nu^{-1}A\right),
\ee
and the solutions  stability  of the variational equation depends on the eigenvalues of the monodromy matrix $\exp\left(
2\pi\nu^{-1}A\right) $.  As a consequence, if we start the integration at $t=0$ from the identity matrix, after a period, we get the relation
$ \exp\left(
2\pi\nu^{-1}A\right)  = Z(2\pi\nu^{-1}) 
$. Thus, this matrix and its eigenvalues can be deduced from a simple numerical integration of the variational equation. If the  eigenvalues modulus  of this matrix are equal to one, the solutions of (\ref{eq:variationH}) are quasi-periodic. Moreover, their fundamental frequencies are $\nu,g_1,g_2$, where $g_1$ and $g_2$ are equal to the eigenvalues arguments of the monodromy matrix multiplied by $\nu/(2\pi)$. 
Fig. \ref{fig:freqH0}  shows the results corresponding to planetary masses $m_1 = m_J$ and $m_2 =  m_S$.
From the numerical computations  of the monodromy matrix  eigenvalues, we conclude that solutions of the variational equation are always quasi-periodic, and thus the invariant manifold of the  quasi-circular orbits $x_1=x_2=0$ is transversally stable, at least in the directions associated to the eccentricities. This property seems to hold for every value of planetary masses that we have tested, that is $m_1= m_2 = 10^{-p}$ and  $m_2= 0.3\times m_1 = 0.3\times10^{-p}$ with $p$ ranging for $3$ to $8$. 
Figure \ref{fig:freqH0} shows the behavior of the frequencies $\nu, g_1$ and $g_2$ along a section of  the space phase. The red curve corresponds to $\nu$, the green one to $g_1$ and the blue one to  $g_2$. The initial conditions are chosen on the segment $\theta = \pi/3$ and $0\leq u \leq u^{(1)}$, $u^{(1)}$ being defined by the relation (\ref{eq:u1}) as the positive intersection of the line $\theta = \pi/3$ with the separatrix $\cS_1$.  This plot shows clearly two different dynamical  domains: the inner one filled with  tadpole orbits ranging from $u=0$ to $u^{(3)}$, and the outer domain for $u^{(3)}<u<u^{(1)}$ populated by horseshoe orbits.   

Inside the inner region, the libration frequency $\nu$ decreases from the value $\nu^0 \approx\sqrt{27(m_1+m_2)/(4m_0)} n_0 \\ \approx 0.0936 \, yr^{-1}$  to zero when the separatrix $\cS_3$ is reached. The frequency $g_1$ associated to the precession of the periastra evolves smoothly between $27(m_1+m_2)n_0/(8m_0)\approx 0.00439 \,  yr^{-1}$ at $L_4$ and $7(m_1+m_2)n_0/(8m_0) \approx 0.00114 \, yr^{-1}$ approaching $\cS_3$.
The box located in the upper left corner of the plot details its evolution for $0<u<0.0055$. As shown in this figure, the last frequency $g_2$ is always very small with respect to the other ones. It starts from zero and reaches zero again at the separatrix, being at least always twenty times smaller than $g_1$. Because in the tadpole region, the frequency $\nu$ is of order $\sqrt{\eps}$ and $g_1$  of order $\eps$, these two frequencies do not generate low order resonances (some of these resonances are indicated by vertical black dotted lines), except in a very narrow neighborhood of $\cS_3$.  As $\nu$ tends to zero at $\cS_3$,  in  both sides of this separatrix, the two curves intersect and $\nu$ becomes smaller than $g_1$. Using the estimates derived by  \cite{Gar1976a,Gar1978a} in the RTBP, one can  easily show that the frequency $\nu$ reaches a logarithmic singularity where it tends to zero as $-(\log\vert u -u^{(3)}\vert)^{-1}$. Consequently, the slope of the curve associated to $\nu$ is very steep and then the low order resonances occur only very close to the separatrix, in a region which is intrinsically unstable.  

The situation is more interesting inside the horseshoe domain. Indeed, if $g_2$ remains always very small with respect to the frequency $g_1$, this one increases significantly after the crossing of the separatrix $\cS_3$. Note that this behavior is already mentioned by  \cite{Morais1999} where the figure 1.a of her paper dedicated to the RTBP corresponds approximately to our figure \ref{fig:freqH0} for $u$ ranging from $0$ to about $0.008$.  After this value, $g_1$ keeps to increase, reaching the resonance $\nu = 2g_1$ at  $u\approx0.01$ and even the 1:1 resonance between $\nu$ and $g_1$ approaching the end of the horseshoe domain materialized by the separatrix $\cS_1$. After the crossing of the two curves representing the frequencies $\nu$ and $g_1$, $g_1$  remains temporarily above $\nu$.    The situation is reversed quickly as $\nu$   tends to zero when $\cS_1$ is reached.

The behavior of the three fundamental frequencies $\nu, g_1, g_2$, described for $m_1= m_J$ and $m_2 = m_S $, seems to be very weakly mass-dependent. Indeed, the simulations performed with the mass sample presented above converge to the same conclusions. First,  the frequency $g_2$ is always small with respect to $g_1$ and of course to  $\nu$. Second, no significant resonance which may destabilize  the average system occurs in the tadpole region,  excepted in a narrow area surrounding  $\cS_3$: the lower the planetary masses are, the larger the ratio $\nu/g_1$ is.  
And third, in the horseshoe domain, low order resonances involving $\nu$ and $g_1$ always occur, in particular, the 1:1 close to  $\cS_1$ is crossed two times.  If these low order resonances generate chaotic behaviors in the average problem, it is not necessarily this mechanism that dominates in the full (non averaged) three-body problem for planetary masses comparable to those of Jupiter or Saturn.   Indeed,  \cite{LauCha2002} deduced from numerical simulations of the planetary three-body problem that horseshoe orbits, even starting with the two planets in circular motion, are unstable for planetary masses satisfying the empirical relation $(m_1+m_2)/(m_0+m_1+m_2) > 0.0004$. This limit corresponding approximately to two Saturn's mass planets  around the Sun.  One can find comparable simulations in different cases in \cite{Dvorak2006}.   In a  nice paper by  \cite{BaOl2006}, this behavior is studied more carefully. In the case of the RTBP, these authors prove that the invariant manifolds associated to $L_3$ deeply penetrate the region populated by horseshoe orbits, generating a large chaotic region, whose size increases with the mass of the secondary (the mass of the primary being fixed). They even mentioned the possible  heteroclinic intersections with the invariant manifolds associated to Lyapunov orbits around $L_1$ and $L_2$. A similar mechanism probably acts in the planar planetary problem. As suggested by C. Sim\'o (private communication), not only $L_3$ invariant manifolds, but also invariant objects of periodic orbits existing in the vicinity of the previous manifolds are supposed to be involved in the process, as it is the case in the spatial RTBP. 
This phenomenon, acting in short time-scale, plays a major role in the instability of the horseshoe regions, even for zero initial eccentricities.
For moderate to small planetary masses, the portion of the horseshoe orbit region intersecting the invariant manifolds mentioned above shrinks to a narrow region,  excluding transitions between the $L_3$ region and the neighborhoods of $L_1$ and $L_2$.  In absence of  short time-scale chaos, the destabilizing effect of the resonances involving the frequencies $\nu$ and $g_j$ can dominate, at least locally, the dynamics of the full problem, as it is the case for the average one.

\section{Beyond the quadratic approximation: Birkhoff's normal form and family of periodic orbits}
 \label{sec:LypFam}

Let us begin with the study of the dynamics in the neighborhood of $L_4$ (the discussion would be the same at $L_5$). 
We first start with the linearized system at this point, or equivalently, with the quadratic expansion of the average Hamiltonian in the vicinity of $L_4$. Using the notations (\ref{eq:dlHamRegCoef}) and (\ref{eq:Hquadh}), this expansion takes the form
\be
\eta_1(\theta-\frac{\pi}{3})^2 + \eta_2 u^2
+ H_2^{(h)},
\ee
the coefficients $\eta_1$ and $\eta_2$  being deduced from (\ref{eq:dlLag}).
A symplectic diagonalization of the associated Hamiltonian system allows one to define a new canonical coordinate system $(z_0,\zt_0,z_1,\zt_1,z_2,\zt_2)$ that reduces the previous quadratic form to 
\be
K_2 = \sum_{q=0}^2 \gam_q z_q\zt_q.
\label{eq:Kdiago}
\ee
As for the variables $x_j$ and $\xt_j$, the coordinates $z_j$ and $\zt_j$ are linked with the relation $\zt_j = -iz_j$.
$L_4$ being an elliptic equilibrium point, the coefficients $\gam_j$ are purely imaginary. More precisely, we have $\gam_0 = i\nu$, $\gam_1 = ig_1$ and $\gam_2 = ig_2 = 0$. Consequently, if $0\leq j, k, l\leq 2$ are three distinct integers, the set defined by the equation $z_j=z_k =0$ is a one-parameter family of periodic orbits of the linearized system parametrized by the  complex number $z_l$. The frequency, which is given by $\vert\gam_l\vert$, is the same for every orbit of the family.  
Let us denote  $\cF_0$ the family parametrized by $z_0$ that corresponds to the quasi-circular motions ($e_1=e_2=0$). The one parametrized by $z_1$  corresponding to the linear approximation of the Anti-Lagrange orbits will be denoted $\cF_1^l$. And the last one,  governed by $z_2$, which  contains  the Lagrangian elliptic configurations, will be symbolized by $\cF_2$.

Let us now  consider the term of degree greater than two in the expansion of the average Hamiltonian in the neighborhood of $L_4$, and let us write this expansion as
\be
K  = K_2 + \sum _{p\geq3} K_{p} \text{ with }  K_p = \sum_{\bfq  \in \cD_{6,p}} \gam_\bfq z_0^{q_0}\zt_0^{\qt_0}z_1^{q_1}\zt_1^{\qt_1}z_2^{q_2}\zt_2^{\qt_2}, 
 \label{coef:K}
\ee 
where
\be
\cD_{2n,p} = \{ \bfq = (q_0, \qt_0, \cdots, q_{n-1}, \qt_{n-1}) \in \NN^{2n}  \slash\, \vert\bfq\vert = \sum_{j=0}^{n-1} 
\left( 
\vert q_j\vert + \vert\qt_j\vert
\right) = p
\}.
\ee 
 All the coefficients $\bfq$ are not allowed in the summations  (\ref{coef:K}): as $\vert x_1\vert^2 + \vert x_2\vert ^2 = \vert z_1\vert^2 + \vert z_2\vert ^2 $ the D'Alembert rule is still valid in coordinates $z_j$, and the non-zero coefficients $\gam_{\bfq}$ verify the relation $q_1 + q_2 = \qt_1 + \qt_2$. 
This last relation imposes that the total degree of the monomials  $z_1^{q_1}\zt_1^{\qt_1}z_2^{q_2}\zt_2^{\qt_2}$ is even, thus, the manifold given by the equation $z_1=z_2=0$ is still invariant by the flow of the Hamiltonian $K$ defined in  (\ref{coef:K}). It turns out that the family $\cF_0$ is not only an invariant set of the linear problem (\ref{eq:Kdiago}) but also of the full average Hamiltonian  (\ref{coef:K}).  
This statement also holds for the family $\cF_2$ including  Lagrange's configurations.  Indeed, as we know that these configurations exist as fixed points of the average problem and that we always have $\theta = \pi/3$ and $a_1 = a_2$ (or $u=0$), $z_0= 0$ along the family $\cF_2$. In addition, the relations $\Delta\varpi = \pi/3$ and $e_1 = e_2 = constant$ impose that $z_1 = 0$, according to Section \ref{sec:L4lin}.
 This  implies additional constraints on the coefficients of the Hamiltonian $K$. Indeed,  as every element of this family is an equilibrium point, the Hamiltonian $K$ fulfills the conditions $ \dron{K}{z_j} =\dron{K}{\zt_j} =0$ when $\vert z_0\vert = \vert z_1\vert =0$. This is equivalent to the cancellation of the coefficients of the terms
\be
(z_2\zt_2)^{q_2}, \, z_0(z_2\zt_2)^{q_2}, \,  \zt_0(z_2\zt_2)^{q_2}, \,  z_1\zt_2(z_2\zt_2)^{q_2}, \, \zt_1z_2(z_2\zt_2)^{q_2}.
\ee

As regards  the Anti-Lagrange family, the relations   $\vert z_0\vert =\vert  z_2\vert =0 $, which characterizes its infinitesimal approximation $\cF_1^l$ does not hold.  Indeed, if these previous relations are preserved by the linear flow of the system associated to $K_2$, it is no more the case by the flow of $K$.  If the Lyapunov center  theorem (see  \cite{MeHa1992})  could be applied to $K$,  it would show the existence of a one parameter family of periodic orbits originating at $L_4$ and tangent to $\cF_1^l$, whose periods would be close to $2\pi/\vert \gam_1\vert$ in the neighborhood of $L_4$. Unfortunately, the coefficient $\gam_2$ being equal to zero, the hypothesis of the latter are not fulfilled.   	  	
To overcome this difficulty, we use a more elaborated method, based on the construction of a Birkhoff normal form.

As mentioned above, the use of the coordinates $(z_j, \zt_j)$, provides an elementary  parametrization of the families $\cF_0$ and $\cF_2$. It is possible to build a coordinate system $(\zeta_j, \zetat_j)$ for which the Anti-Lagrange family possesses the same kind of parametrization than  the two other families, that is $\vert \zeta_0 \vert = \vert \zeta_2\vert = 0$ and $\zeta_1$ depending on the element of the family. This coordinate system can be chosen among one of those that reduce the Hamiltonian $K$   to its Birkhoff's normal form.
In this context, the Birkhoff transformation consists in the construction of  canonical transformations that act on homogeneous polynomials of given degrees in order to eliminate  non-resonant monomials. These are the monomials which are not of the form
\be
z_0^{q_0}\zt_0^{q_0}z_1^{q_1}\zt_1^{q_1}z_2^{q_2}\zt_2^{q_2}.
\ee
More precisely, this transformation is performed iteratively, each step being dedicated to the normalization of a given degree. 
An elementary transformation is defined by the time-one map of the flow of an auxiliary Hamiltonian $w_n$ defined as a solution of the equation 
\be
 \gam_0\left( \zetat_0\dron{w_n}{\zetat_0} - \zeta_0\dron{w_n}{\zeta_0} \right)
+
\gam_1\left( \zetat_1\dron{w_n}{\zetat_1} - \zeta_1\dron{w_n}{\zeta_1} \right)
= \Psi_n,                                                                                                                             
\label{eq:homolo}
\ee
where $\Psi_n$ contains  non resonant monomials of degree $n$ (see \cite{Morbidelli02}).
This equation being linear, it can be solved monomial by monomial. The resolution of the equation (\ref{eq:homolo}) for $\Psi_n =  z_0^{q_0}\zt_0^{\qt_0}z_1^{q_1}\zt_1^{\qt_1}z_2^{q_2}\zt_2^{\qt_2}$ introduces the divisor 
$ \gam_0(\qt_0-q_0) + \gam_1(\qt_1-q_1)$.
If we assume that $\gam_0$ and $\gam_1$ are rationally independent, which is generically the case\footnote{As $\gam_1/\gam_0 \sim \sqrt{27(m_1+m_2)/m_0}/4$, only high order resonances can occur. This allows one to build the normal form up to a high degree, typically of order $1/\sqrt{\eps}$. In our numerical application, the first potential small denominator involves terms of degree $48$.}, the denominator cancels only if $q_0=\qt_0$ and $q_1=\qt_1$ independently of $q_2$ and $\qt_2$. Using the D'Alembert rule, the only monomials involving divisors equal to zero are $z_0^{q_0}\zt_0^{q_0}z_1^{q_1}\zt_1^{q_1}z_2^{q_2}\zt_2^{q_2}$ which are not eliminated from the Hamiltonian. Consequently, the Birkhoff normal form can be computed at any degree.  
Let us denoted by $(\zeta_j,\zetat_j)$ the normalizing coordinates. By construction, the coordinates $(z_j,\zt_j)$ and $(\zeta_j,\zetat_j)$  are related by expressions of the form: $z_j = \zeta_j + \gO_2(\zeta_j,\zetat_j)$ with $ \zetat_j = -i \zetab_j$. Then the Hamiltonian reduced to a Birkhoff normal form reads
\be
N(\zeta_j,\zetat_j) =  \sum_{q=0}^2 \gam_q\zeta_p\zetat_p + \sum_{q_0+q_1+q_2\geq 2} \gam'_{\bfq}(\zeta_0\zetat_0)^{q_0}(\zeta_1\zetat_1)^{q_1}(\zeta_2\zetat_2)^{q_2},
\ee
where the $\gam'_{\bfq}$ are complex  numbers  such that the coefficients of the monomials $(\zeta_2\zetat_2)^{q_2}$ vanish. 
As an example, the Birkhoff normal form corresponding to $m_1 = m_J$ and $m_2 = m_S$ computed up to the fourth degree in $\zeta_j, \zetat_j$ reads
\be
\begin{split}
             -0.093622\, i \zeta_0\zetat_0 
             - 0.00439\, i\zeta_1\zetat_1  
              -       2450.55\, \zeta_0^2\zetat_0^2
 +       472.218\, \zeta_0\zetat_0\zeta_1\zetat_1 \\
  +       253.10\, \zeta_0\zetat_0\zeta_2\zetat_2
  -        38.0734\, \zeta_1^2\zeta_1^2
-        1.17035\, \zeta_1\zetat_1\zeta_2\zetat_2,
  \end{split}
 \ee
 where only a few  digits of the coefficients are given here.  Remark that the ``linear" fundamental frequencies, namely the coefficients of the monomials  $i \zeta_0\zetat_0$ and $i \zeta_1\zetat_1$, are negative real numbers. As it is more convenient to deal with positive quantities, we have decided to change the sign of these frequencies in the previous sections.  
In the coordinates $(\zeta_j, \zetat_j)$, the Hamiltonian system associated to $N$ is trivially integrable. In particular, its phase space is foliated in 3-dimensional invariant tori carrying linear flows. In other words, using the angle-action variables $(\varphi_j,I_j)$ defined by the relations $\zeta_j = \sqrt{I_j}e^{i\varphi_j}$, one can verify that the actions $I_j$ are integrals of the motion, and that every solution is quasi-periodic with fundamental frequencies equal to $\om_j = \dron{N}{I_j}$.  
Among these solutions, we focus now on those that are members of the families $\cF_j$, that is the solutions satisfying the relations  $I_k = I_l =0$ ($j, k, l$ pairwise distinct), or equivalently $\zeta_k = \zeta_l=0$.
Using the transformation that reduces the Hamiltonian $K$ to its normal form up to a given degree, say $2n$, and taking into account the symmetries of the transformation\footnote{It is not necessary to detail this transformation, but the key point lies on the fact that it takes the form $\zeta_j = z_j + f_j(z_1,z_2,z_3,\zt_1,\zt_2,\zt_3)$ where the polynomial $f_j$ possesses the same symmetries as $\dron{K}{\zt_j }$.},
one can show that the families are parametrized as follows.
The family $\cF_0$ containing the quasi-circular periodic orbits is given by
\be
 z_0 = \zeta_0 + f(\zeta_0,\zetat_0),\, z_1 = z_2 =0, \, \zetat_0 = -i\zetab_0 \in \mathbb{C},
\ee
$f(\zeta_0,\zetat_0)$ being a polynomial of degree $2n$ in $(\zeta_0,\zetat_0)$ whose lower order terms are quadratic.
The family $\cF_1$ associated to the Anti-Lagrange orbits reads
\be
z_0 = P(\zeta_1\zetat_1), \,
z_1 =\zeta_1 + \zeta_1 Q(\zeta_1\zetat_1), \,
z_2 = \zeta_1R(\zeta_1\zetat_1), 
\label{eq:paramF1}
\ee
where $P, Q$ and $R$ are polynomials of a single complex variable of degree $n$ whose lower order term is of degree one. Of course, we still have $\zetat_1 = -i\zetab_1 \in \mathbb{C}$. 
As mentioned above, the elliptic equilateral configurations $\cF_2$ are still given by:
\be
 z_0 = z_1 = 0, \,  z_2 = \zeta_2, \, \zeta_2 = -i\zetab_2 \in \mathbb{C} \, .
\ee
%
%
\begin{figure}[h!]
$$\begin{array}{rl}
\hspace{-1.3cm}\includegraphics[width=8cm]{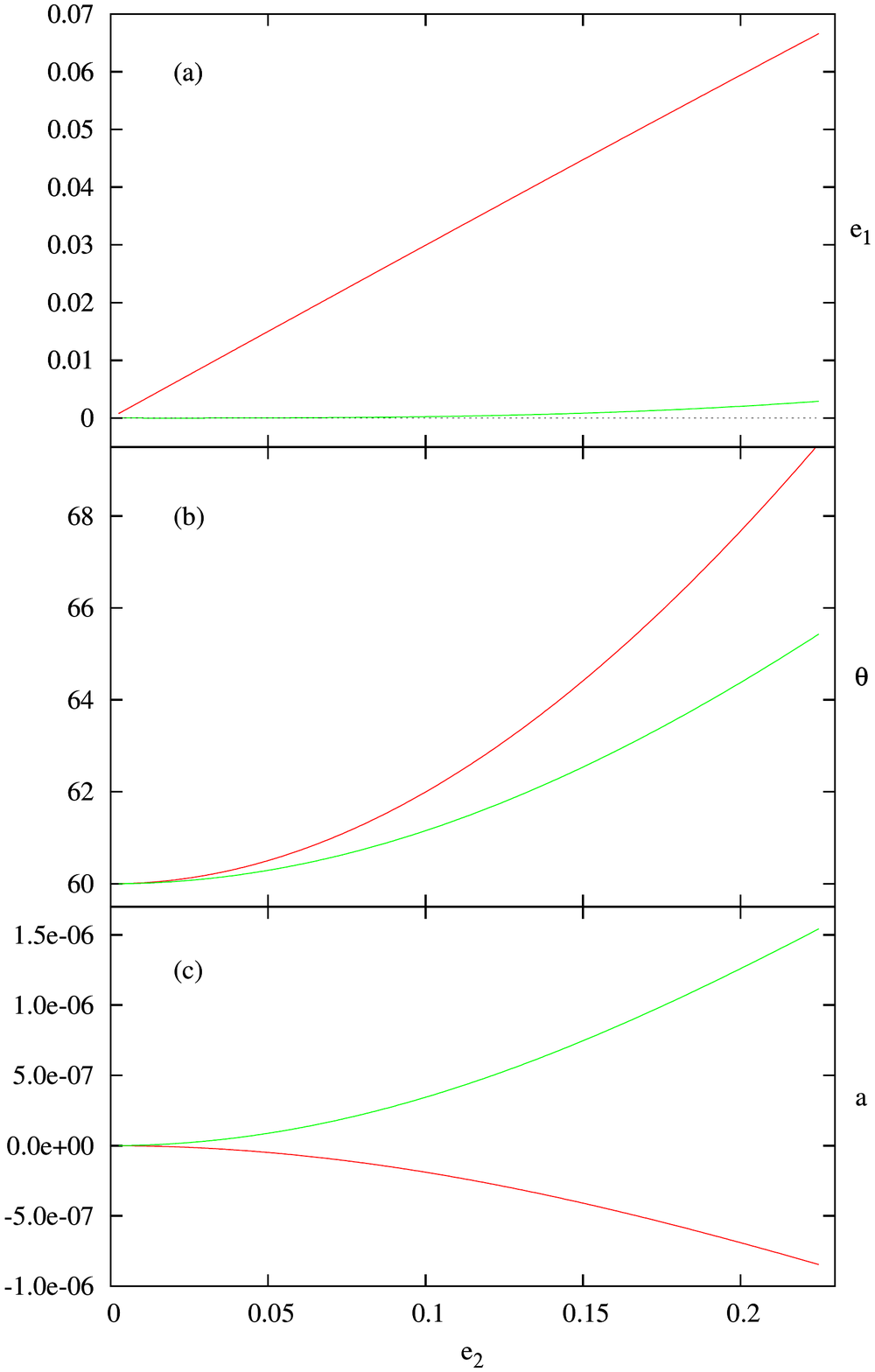} &
\hspace{-2cm}\includegraphics[width=8cm]{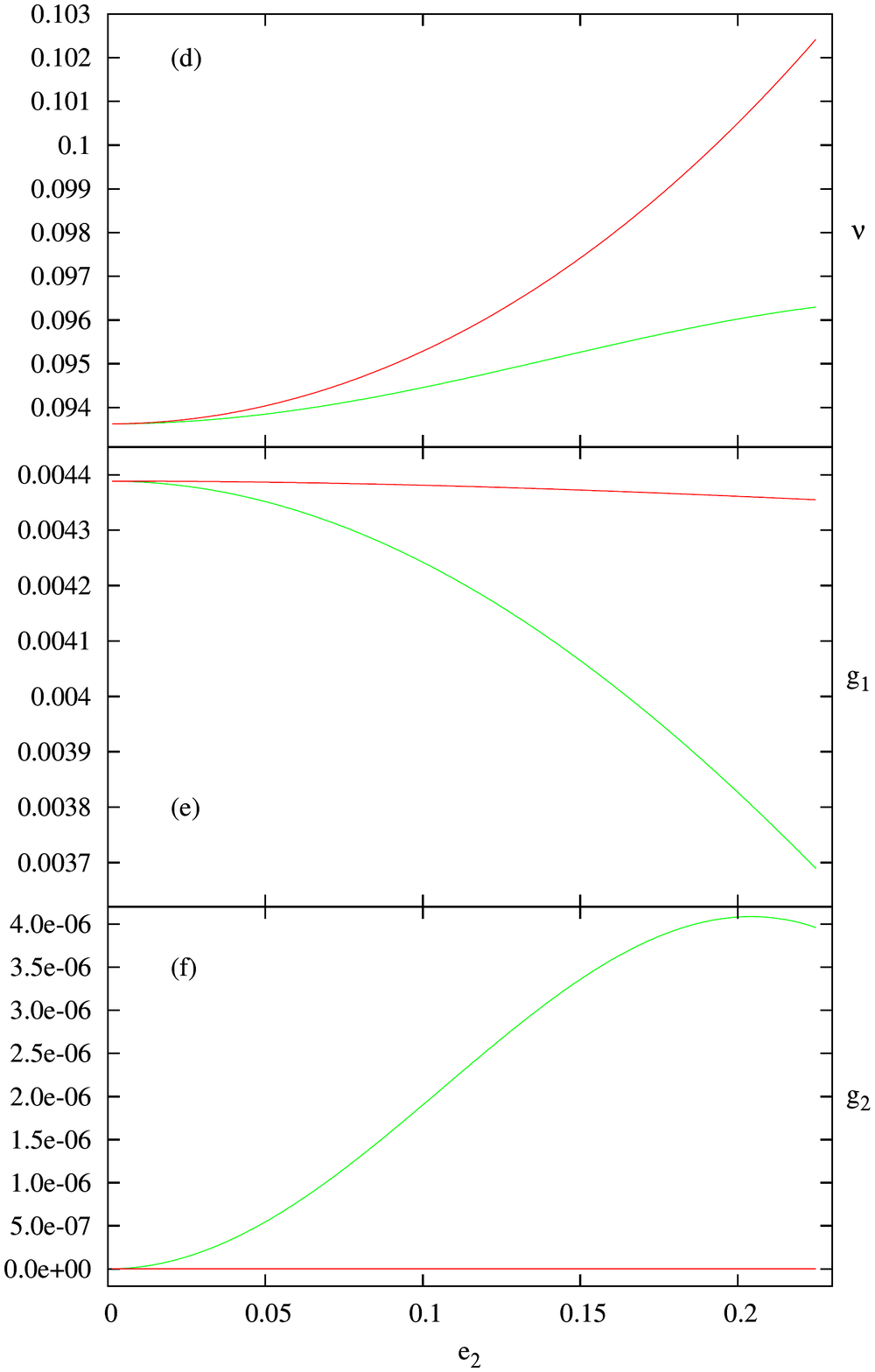} 
\end{array}$$
\caption{Evolution of the elliptic elements and of the fundamental frequencies in function of $e_2$, along the families $\cF_1$ and $\cF_2$. Left panel: elliptic elements. $e_1$ red curve, panel (a). $\theta$ in red and $\Delta\varpi$ (green) in panel (b). Panel (c): $a_1-1$ (red), $a_2-1$ (green).  Right panel: the fundamental frequencies $\nu$, $g_1$ and $g_2$ are represented in (d), (e) and (f). The green curves correspond to frequencies computed  along $\cF_1$, $\cF_2$ in red.  }
\label{fig:Birkhoff}
\end{figure}
$\cF_1$ is the most interesting of these three families. Indeed, the quasi-circular family $\cF_0$ is well known and its orbits are already represented in the  figures  \ref{fig:figH0} and \ref{fig:figCompar}. The elliptic equilateral configurations $\cF_2$  has also been extensively studied since their discovery by Lagrange. In addition, its expression in terms of elliptic elements is well known since it corresponds to $a_1=a_2$, $e_1=e_2$ and $\theta = \Delta\varpi = \pi/3$  (or $-\pi/3$ for the family starting from $L_5$).
On the contrary, the family $\cF_1$ has been only partially studied by   \cite{GiuBeMiFe2010}  and  \cite{HaVo2011}. 

The use of a Birkhoff normal form allows one to get any desired information concerning this family, providing that the orbits of the  family are contained inside the domain of validity of the normal form. Practically, this is not the case for the whole family, at least some portion of $\cF_1$ including $L_4$ is contained in such a domain. In order to estimate this region, we have computed the relative difference between $K$ and $N$ along $\cF_1$ using the expression
\be
\rho(\zeta_1) = \frac{\vert K(z_0,z_1,z_2,\zt_0,\zt_1,\zt_2) - N(0,\zeta_1,0,0,\zetat_1,0)\vert}{\vert N(0,\zeta_1,0,0,\zetat_1,0)\vert},
\ee
where the values of $z_j$ are deduced  from $\zeta_1$ by the relations (\ref{eq:paramF1}).  As $\rho(0) = 0$, and in order to be consistent with the approximations done during the computation of the average Hamiltonian, we consider that the normal form is relevant while $\rho(\zeta_1) < \eps^2$.  We have estimated that the  thirtieth degree  was a good compromise between the precision of the normal form and its number of terms. Once  defined this domain in which the normal form is relevant, a linear transformation allows one to express the $z_j$ (deduced from $\zeta_j$) in terms of $\theta, u, x_1, x_2$ and to deduce the expression of $\cF_1$ with the help of the elliptic elements.  This is shown on the left panel of the figure \ref{fig:Birkhoff} in the particular cases $m_1 = m_J, m_2 = m_S$.   Fig. \ref{fig:Birkhoff}.a displays the evolution of the eccentricity $e_1$ versus $e_2$ along the family $\cF_1$ (red curve). The maximal value of $e_2$ for which the condition  $\rho(\zeta_1) < \eps^2$ (here $\rho(\zeta_1) <10^{-6}$)  is fulfilled is  $e_2 = 0.23$, which corresponds to $e_1 \approx 0.066$. Let us note that, we have $\rho(\zeta_1) <3\times 10^{-16}$ as long as $e_2<0.12$ and that the precision obtained using the Birkhoff normal form is comparable to the machine epsilon. 
In this domain, $e_1$ seems to depend linearly on $e_2$, the slope of the (red) line being equal to $m_2/m_1$.  The difference between the green curve, which shows the variation of  $m_1e_1 - m_2e_2$ versus $e_2$, and the dashed black line ($e_1 = 0$), indicates that the relation $m_1e_1 = m_2e_2$ is fulfilled only at the origin of the family $\cF_1$.  \cite{HaVo2011} suggest that along this family, $e_1$ and $e_2$ tend simultaneously to one regardless of the planetary masses. 
 Fig. \ref{fig:Birkhoff}.b  shows how the angles $\theta$ in red and $\Delta\varpi - 180^{\circ}$ in green move away from their value at the origin when  $e_2$ increases. Basing on numerical simulations, \cite{HaVo2011}  suggest that the angles $\theta$ and $\Delta\varpi$ tend  to $180^{\circ}$, when the eccentricities tend to one, which would correspond to a triple collision. 
 The last figure of the left panel,   Fig. \ref{fig:Birkhoff}.c, shows the slight deviation of the semi-major axes from the equality $a_1=a_2=1$. Practically, $a_1 -1$ is plotted in red, while the green curve corresponds to $a_2-1$. This figure shows that, at least for $e_2<0.23$, the variations of the semi-major axes  are very small (of order $\eps^2$) compared  to the other elliptic elements. The situation may be different for large values of the eccentricities, but this is not mentioned in the literature. 
 
  Remark that, with the help of the analytical expression of $\cF_1$,  we analyze a relatively small portion of the family $\cF_1$ compared to the region studied numerically in \cite{GiuBeMiFe2010} and  \cite{HaVo2011} where the eccentricities reach  $0.8$.
In contrast, our analytical study allows us to access to more information. 
First, it provides a complete understanding of the dynamics of all quasi-periodic trajectories lying in the validity domain of the Birkhoff normal form. Second, using an analytical expansion of the eigenvectors of the differential system (\ref{eq:variationH}), we can establish rigorously  that, at the beginning of the family $\cF_1$, the orbits satisfy the relation $m_1e_1 = m_2e_2$, which has been empirically deduced from numerical simulations in  \cite{GiuBeMiFe2010}. Third, it allows us to compute straightforwardly the fundamental frequencies associated to each trajectory belonging to a given family. Indeed, for  $\cF_l$, the derivative of the normal form $N$ with respect to $I_l$ is the frequency of the corresponding periodic orbit of the family (this frequency is zero in the particular case of $\cF_2$). The normal frequencies are obtained by derivation with respect to the two other action variables.  These three frequencies are plotted in Fig.\ref{fig:Birkhoff}.d-f for the families $\cF_1$ and $\cF_2$. The fundamental frequencies associated to the family $\cF_0$ are not represented here for the simple reason that the normal form furnishes the same values as in figure \ref{fig:freqH0}, at least in a neighborhood of the circular equilateral configuration $L_4$. 
The frequency $\nu$ (resp. $g_1$, $g_2$) is plotted in Fig. \ref{fig:Birkhoff}-d (reps. \ref{fig:Birkhoff}-e, \ref{fig:Birkhoff}-f). The red curves correspond the equilateral family $\cF_2$ while the green curves are associated to $\cF_1$. 

Although these frequencies and their derivatives are equal at the origin of the families, their behaviors along $\cF_1$ and $\cF_2$ are very different.
As shown  Fig. \ref{fig:Birkhoff}-f, the frequency $g_2$ is obviously equal to zero all along the Lagrange family since these trajectories are fixed points of the average problem. On the contrary, computed on the family $\cF_1$  this frequency increases to a (local) maximum  although it remains small in the considered interval.  According Fig. \ref{fig:Birkhoff}-e, $g_1$  changes only very slightly for the equilateral family, but very much for $\cF_1$. 
Remark that the quantity $2\pi/g_1$, which seems to increase with the distance to $L_4$, is the period of the orbits belonging to $\cF_1$.  Regarding $\nu$ (Fig. \ref{fig:Birkhoff}-d), the frequency associated to $\cF_1$ seems to reach a local maximum, while the one corresponding  to $\cF_2$ increases.  

What can be said concerning the behavior of the fundamental frequencies outside of the validity domain of the normal form? One thing is clear about the equilateral configurations: when their eccentricity increases, a critical value depending on the mass ratio $(m_0m_1+m_0m_2+m_1m_2)/(m_0+m_1+m_2)^2$ is reached,  leading to a period-doubling bifurcation where the family looses its stability  \citep{Robe2002, Nauenberg2002}. Consequently, for $e_2>0.23$,  the frequency $\nu$ is supposed to keep  increasing, until it reaches the resonance $2\nu = n$, where $n$ is the planetary mean motion (close to one if $\ab = \cG=m_0=1$).  This is certainly the mechanism that was acting when  \cite{GiuBeMiFe2010} observed the shrinking of   the stable region surrounding the equilateral equilibrium,  and finally its fading  when  the eccentricity grows.
The way that the family $\cF_1$ ends  is less clear.  In fact, at high eccentricities, only numerical simulations of these orbits have been performed \citep{ GiuBeMiFe2010, HaVo2011}, and when $e$ does not exceed $0.8$.  \cite{HaVo2011} suggest that for high eccentric orbits, the two eccentricities coincide, and that $\theta$ and $\Delta\varpi$ tend to $\pi$. 
This would imply that the Anti-Lagrange family $\cF_1$, and the Euler family originating at $L_3$ intersect, or end  at a triple collision. 
This conjecture has to be checked.

\section{Concluding remarks}
\label{sec:comments}

In this paper, we developed a Hamiltonian formalism adapted to study the motion of two planets in co-orbital resonance.  This analytical formalism intends to unify several works dedicated to the 1:1 mean-motion resonances like the formulations developed by  \cite{Ed1977} or 
\cite{Morais1999,Morais2001} in the case of the RTBP, but also models obtained by \cite{DeMu1981a} and \cite{YoCoSyYo1983} aiming to understand the dynamics of the two Saturn's satellites Janus and Epimetheus.  

Our approach consists on an expansion of the average Hamiltonian in power series of both planetary eccentricities and inclinations. To make the study of the tadpole orbits  as well as the horseshoe orbits possible, an expression of the mutual distance valid for all values of $\theta= \lam_1 - \lam_2$ has been introduced in the Hamiltonian. 
Contrary to the other authors who modeled the distance between the two planets by the term $\sqrt{2-2\cos\theta}$, we have chosen to introduce the divisor $\sqrt{a_1^2+a_2^2 -2a_1a_2\cos\theta}$.  This changes drastically the topology of the integrable problem associated  to $e_1=e_2=I_1=I_2=0$. Indeed, the usual model, which possesses  three fixed points corresponding to $L_3$, $L_4$ and $L_5$, is singular when $\theta=0$, regardless of the  planetary semi-major axes values. 
Our approximation gives rise to two additional fixed points corresponding to the Euler points $L_1$ and $L_2$. The singularity, that is usually identified to a line in the usual model, is here reduced to a single point that corresponds to the collision of the two planets in the same circular orbit, that is  $a_1=a_2, \theta=0$.
Thus, the topology of the two problems is very different. Indeed, with the first approximation, the phase space is divided in three distinct regions: two symmetrical  libration regions around $L_4$ and $L_5$  respectively, and a third one, populated with horseshoe orbits that encompass  the three equilibrium. Inside this last region, the semi-major axes tend to infinity when the angle $\theta$ approaches zero, which is obviously not very realistic.  With the average model presented in this paper, the two regions surrounding $L_4$ and $L_5$ are practically the same as in the usual model, while the horseshoe region bounded in a domain lying between  the separatrix emanating from $L_3$ and the one originated from $L_2$.  This model can be useful to simulate  captures or  transitions between different kinds of trajectories under the influence of  weak dissipations, or slow migrations. Indeed, contrary to the usual model, the non-resonant region is better separated than the resonant horseshoe region.

For small eccentricities, the global topology of the problem is similar to the one described in  \cite{NeThoFeMo02} in the RTBP framework, using numerical averaging methods which are not limited to moderate eccentricities and inclinations. Although we are constrained by the size of eccentricities and inclinations, our model possesses  at least two advantages. On the one hand, this average   problem, as long as the number of terms of its Hamiltonian is not too large, allows fast numerical simulations using large time-steps.
On the other hand,    the present analytical formulation of the problem can help to obtain theoretical results concerning the stability inside the co-orbital resonance. If much has been done in the vicinity of the equilateral equilibrium points, especially in the RTBP (see  \cite{GaJoLo2005} and references therein), the theoretical stability of horseshoe orbits  remains an open problem. 

With the help of a Birkhoff normal form, we have shown how  the equilateral family $\cF_2$ and the Anti-Lagrange family $\cF_1$ bifurcate from the circular equilateral configuration $L_4$. If the behavior of the family $\cF_2$ is well known from its beginning at $L_4$ to its termination by a period-doubling bifurcation \citep{Robe2002}, the same cannot be said for the family $\cF_1$. 
At this point, we only have conjectures concerning the termination of this  family. This might be a triple collision, and could be related to the end of the Euler aligned configurations originated at $L_3$.  A similar question, which is not discussed in the present paper, concerns the so-called quasi-satellites family (see   \cite{HaPsyVo2009,GiuBeMiFe2010})  which could also end by collisions when the eccentricities tend to one (an alternation of two kinds of double collisions involving on the one hand, Sun and a first planet, and on the other hand, the second planet and the Sun). 

A last point should be mentioned. In Section \ref{sec:fixedpoints}, the vertical variational equation has been set aside because the quadratic part of Hamiltonian in inclination was equal to zero.  A careful study of this situation would reveal interesting bifurcation phenomena giving rise to families of remarkable orbits, as in the case of the RTBP \citep{PeZa1991,Merchal2009} or in the general three-body problem with equal masses \citep{FeChe2008}.
Finally, a lot remains to be done in that field.

\section{Appendix: $L_4$ in heliocentric canonical elliptic elements}
\label{sec:appendix}

Let us assume that the three bodies describe a circular Lagrangian equilateral configuration where $\rho$  is the length of the triangle sides. The heliocentric coordinate system can be chosen such as $\br_j= \rho \bu_j$  where 
\be
\bu_j = \vect{\cos\varphi_j}{\sin\varphi_j }{0}, \quad \text{with}\, 
\varphi_1 = \om t \,\text{and}\,  \varphi_2 = \om t +\dfrac{\pi}{3},
\ee
the angular velocity $\omega$ of the relative equilibrium satisfying the third Kepler law  $\omega^2\rho^3 = \mu = \cG(m_0+m_1+m_2)$. 
The elliptic elements $(a_j,e_j,v_j,\varpi_j)$ can be  derived from the  canonical heliocentric coordinates $(\brt_j,\br_j)$ using the relations
\be
K_j =   \frac{\brt_j^2}{2\beta_j} - \frac{\mu_j \beta_j}{\norm{\br_j}} = -\frac{\beta_j\mu_j}{2a_j},
 \ee
 \be 
\bE_j = \mu_j^{-1}\frac{\brt_j }{\beta_j}\times \left( \br_j\times\frac{\brt_j }{\beta_j} \right) - \bu_j = Êe_j\vect{\cos\varpi_j}{\sin\varpi_j }{0}
\ee
and 
\be
\cos v_j =  e_j^{-1} \bE_j\cdot \bu_j.
\label{eq:cos}
\ee
\be
\frac{\brt_j}{\beta_j} = \gam^{-1}\left( \dot\br_j -\frac{\beta_k}{m_0} \dot\br_k \right),
 \, \text{with} \quad \gam = 1- \frac{\beta_1\beta_2}{m_0^2},
 \quad (j,k)\in\{1,2\} \quad\text{and} \quad j\ne k,
 \label{eq:moment_vitesse}
\ee
a straightforward computation leads to the expressions
\be 
K_j = -\frac{\beta_j\mu_j}{2\rho}
\left(
2 - \frac{\mu}{\mu_j} \left( 1 - \frac{\beta_k}{m_0} + \frac{\beta_k^2}{m_0^2}\right)\gam^{-2}
\right)
\label{eq:KeplerEllipt}
\ee
and
\be
 \gam^2 \bE_j = \left( 
\frac{m_k}{m_0+m_j} -\frac12\frac{\mu}{\mu_j}\frac{\beta_k}{m_0}
\right)\bu_j + 
\frac{\mu}{\mu_j}\frac{\beta_k}{m_0}\left( 
\frac12\frac{\beta_k}{m_0} -1
 \right) \bu_k.
\label{eq:EccEllipt}
\ee 
According to (\ref{eq:KeplerEllipt}), the semi-major axis of the planet $j$ is a time-independent quantity approximated by the expression
\be
a_j = \rho\left( 
1 +\frac{m_k}{m_0} \frac{m_1+m_2}{m_0}  +\gO(\eps^3) 
                           \right)
\ee
which  is slightly larger than the radius $\rho$ of the configuration. 
As $\bu_1\cdot\bu_2 = 1/2$, the expression (\ref{eq:EccEllipt}) shows that the eccentricity (modulus of $\bE_j$) is constant, and that the ellipse rotates with an angular velocity equal to $\omega$.
A first order expansion of (\ref{eq:EccEllipt}) gives 

\be 
\bE_j = \frac{m_k}{m_0}\left(  \frac{\bu_j}{2} - \bu_k \right) +\gO(\eps^2) 
\ee
and 
\be
e_j =  \frac{\sqrt3}{2}\frac{m_k}{m_0} +\gO(\eps^2). 
\ee
We deduced from  (\ref{eq:cos})  that the true anomalies $v_j$ of the planets  satisfy
\be
\cos v_j =    \frac{ 4m_j + m_k}{2\sqrt{3}m_0} +\gO(\eps^2). 
\ee


\newcommand{\noopsort}[1]{}

\end{document}